\documentclass[aps,pre,twocolumn,floatfix,longbibliography,showpacs,superscriptaddress,groupedaddress]{revtex4-1}  
\usepackage{graphicx}
\usepackage{amssymb}
\usepackage[utf8]{inputenc}
\usepackage{amsmath}
\usepackage{tikz}
\DeclareGraphicsExtensions{.pdf}
\graphicspath{{./prefigs/}}
\usepackage{bigints}
\usepackage{hyperref}

\let\at@
\catcode`@=\active
\def@#1{\ifmmode\boldsymbol{#1}\else\at#1\fi}
\let\quot"
\catcode`"=\active
\def"#1"{``#1''}

\begin{document}

\title{Stochastic Bifurcations in the Nonlinear Parallel Ising Model}
\author{Franco Bagnoli}
\affiliation{Dipartimento di Fisica e Astronomia and CSDC, Universit\`a di Firenze,
 Via G. Sansone 1, I--50019 Sesto Fiorentino, Italy, 
 also INFN, sez.\ Firenze. email:franco.bagnoli@unifi.it}

\author{Ra\'ul Rechtman}
\affiliation{Instituto de Energías Renovables, Universidad Nacional Autónoma de México
Apdo. Postal 34, 62580 Temixco, Mor., Mexico. email:rrs@ier.unam.mx}

\begin{abstract}
We investigate the phase transitions of a nonlinear, parallel version  of the Ising model, characterized by an antiferromagnetic linear coupling  and ferromagnetic  nonlinear one.  This model arises in problems of opinion formation. The mean-field approximation shows chaotic oscillations, by changing the couplings  or the connectivity.
The spatial model shows bifurcations in the average magnetization, similar to what seen in the mean-field approximation, induced by the change of the topology, after rewiring short-range to long-range connection, as predicted by the small-world effect. These coherent  periodic and chaotic oscillations of the magnetization reflect a certain degree of synchronization of the spins, induced by long-range couplings.  
Similar bifurcations may be induced in the randomly connected model by changing the couplings or the connectivity and also the dilution (degree of asynchronism) of the updating. 
We also examined the effects of inhomogeneity, mixing ferromagnetic and antiferromagnetic coupling, which  induces an unexpected bifurcation diagram with a "bubbling" behavior, as also happens for dilution. 
 \end{abstract}
\pacs{05.45.Ac,05.50.+q,64.60.aq,64.60.Ht}

\maketitle

\section{Introduction}

There are quite a large number of studies about opinion formation in uniform societies~\cite{Hegselmann,Deffuant,review,Stauffer, GalamReview,Galam1, guazzini,BagnoliGuazziniLio,GrottoGuazziniBagnoli,GuazziniCiniBagnoliRamasco,ViloneCarlettibagnoliGuazzini}.
Many such models adopt an approach similar to that of the Ising model. In such cases one has
two opinions, say A and B or -1 and 1, and one is interested in the establishment of a 
majority (magnetic phase transitions) or in the effects of borders, or in the influence of some leader (social impact theory)~\cite{latane}. 
This opinion space can be seen as the first ingredient of these models. 

The second ingredient is how to model the response to an external influence. It is common to
classify the attitude of people (agents) as either conformist or
contrarian (also known as nonconformist). 
A conformist tends to agree
with his neighbors and a contrarian to disagree. 

It is also easy to map this attitude onto Ising terms:   conformist agents correspond to 
 ferromagnetic coupling and contrarians to
antiferromagnetic ones~\cite{review}.
The effects induced by the presence of contrarian agents in a society have been studied in models
related to the voter model~\cite{masuda2013,crokidakis2014,Independence, schneider2004,
delalama2005,corcos02,galam04,Biswas,Galam-Gemrev, Galam-chaotic, sudoyi2013, bagnoli2013, bagnoli2015}.

In general, agents that are under a strong social pressure tend to agree with
the great majority even when they are certain that the majority's
opinion is wrong, as shown by Asch~\cite{Asch}.   Under a strong social
pressure a contrarian may agree with a large majority, an phenomenon that may be modelled using non-linear interactions.  

The strategy of following an overwhelming majority may be ecological, since it is probable that this coherent behavior
is due to some unknown piece of information, and in any case the
competitive loss is minimal since it equally affects all other agents.

A binary opinion model where an agent tends to align with the largest neighboring cluster, similar to an Ising model with plaquette interactions, was studied in Ref.~\cite{biswas2009}. In this model, a single dissenting agent immersed in a cluster of different opinions cannot overcome the social pressure, and therefore the model exhibits absorbing homogeneous phases. The possibility of dissenting, distributed as a quenched disorder, was introduced in Ref~\cite{biswas2011a}. 
 
The third ingredient is the connectivity, i.e., how the neighborhood of a given agent is composed. 
Traditionally, magnetic systems have been studied either on regular lattices, trees or with random connections, whose behavior is 
similar to that of the mean-field approach. In recent years, much attention  has been devoted to other network topologies, 
like the small-world~\cite{WattsStrogatz} to scale-free~\cite{Barabasi} etc.

The Ising model has been studied in different topologies~\cite{Klemm,WuZhou,HolmeNewman,barre}, in particular, the topological details may affect the critical dynamics~\cite{goswami2011} and the zero-temperature quench~\cite{biswas2011}.

In contrast with the Ising model, in the study of opinion formation there is no compulsory obligation to have symmetric interactions, each agent is influenced by those in his neighborhood, which are not necessarily influenced by the first agent.  

Each individual may be a conformist
or a contrarian and this character does not change in time. In these terms, the simple
ferromagnetic Ising model represents a uniform society of conformists
with local symmetric interactions.

The fourth ingredient is the update scheduling, that may be completely asynchronous, 
like in standard Monte Carlo simulations, or completely parallel, 
like in Cellular Automata, or something in between~\cite{Derrida,NewmannDerrida,Cirillo}. It is not clear which scheme is the most representative of reality. 
Real human interactions 
are indeed continuous, but also clocked by days, elections, etc. 

An effect that is favoured by parallelism is synchronization in the presence 
of complex dynamics. As happens in physics, a macroscopic irregular behavior (macroscopic chaos) 
implies a coherent, although irregular motion of many elements (the microscopic constituents). 

One of most intriguing effects is the hipster's one, in which a
society of contrarians tends to behave in a uniform way~\cite{hipster}. Clearly, ``conformist
hipsters'' always change their behavior, when they realize to be still
in the mainstream, but since they do so all together, they remain synchronized: the parallelism is a crucial element of 
such a behavior. 

Finally, the fifth ingredient is homogeneity. There are many possibilities of introducing mixtures of agents or spins with different coupling. We investigate what happens when one mixes ferromagnetic and antiferromagnetic interactions, and we shall show that this mixture promotes a ``bubbling'' behavior in the bifurcation, meaning that the bifurcation appears first for intermediate values of the parameters, similar to what happens with asynchronism.

In previous studies we presented ``reasonable contrarian'' agents whose response to the average opinion of their neighbors is nonlinear and discusssed the collective behavior of societies composed of reasonable contrarians only and by mixtures of these agents and nonlinear conformists~\cite{bagnoli2013, bagnoli2015}. The rationale was that in some cases, and in
particular in the presence of frustrated situations like in minority
games~\cite{minority,minority1,minority2,minority3}, it is not convenient to always follow the
majority, since in this case one is always on the ``losing side'' of
the market.  This is one of the main reasons for the emergence of a
contrarian attitude. On the other hand, if all or almost all agents in
a market take the same decision, it is often wise to follow such a
trend. We can denote such a situation with the word ``social norm''.

A society  composed by a strong majority of reasonable contrarians exhibits
interesting behaviors when changing the topology of the
connections. On a one-dimensional regular lattice,
there is no long-range order, the evolution is disordered and the
average opinion is always halfway between the extreme values~\cite{bagnoli2005}. However, 
adding long-range connections or rewiring existing ones,
we observe the Watts-Strogatz ``small-world'' effect, with a
transition towards a mean-field behavior.  But since in this case the
mean-field equation is, for a suitable choice of parameters, chaotic,
we observe the emergence of coherent oscillations, with a bifurcation
cascade eventually leading to a chaotic-like behavior of the average
opinion.  

The small-world transition is essentially a synchronization
effect.  Similar effects with a bifurcation diagram resembling that of
the logistic map have been observed in a different model of ``adapt if
novel - drop if ubiquitous'' behavior, upon changing the
connectivity~\cite{Dodds, Harris}.

The main goal of the present study is that of reformulating the opinion formation models mentioned above~\cite{bagnoli2013, bagnoli2015}, in terms of a parallel, nonlinear Ising model both on a
regular lattice, where the spin at any site is influenced by its
nearest neighbors, and on small-world networks.

In the first case
the mean-field behavior of the magnetization is described by a
nonlinear equation for which chaos can be evaluated by the Lyapunov
exponent~\cite{ott02},
which is a measure of the stability of trajectories. 

The Lyapunov exponent is the time average of the growth rate of an initial infinitesimal perturbation of a 
trajectory. Clearly, this quantity cannot be simply defined for stochastic systems, since in this 
case one would essentially measure  the effects of the noise. However, in many cases and in particular the present one, we would like to compare the dynamical properties 
of a stochastic microscopic model and its mean-field approximation.
We show here that the Boltzmann's entropy of an aggregate variable like the magnetization is 
a quantity that can be defined for both deterministic and stochastic systems. In the first case, Boltzmann's entropy can be used as a measure of chaos~\cite{Boltzmann}.

The scheme of the paper is the following. We discuss the ``nonlinear'' parallel Ising model in Section~\ref{sec:parallelising}. 
We can therefore introduce the mean-field approximation of the model in Section~\ref{sec:meanfield}, showing the bifurcation 
phase diagrams as a function of the parameters. 
The definition of the entropy and the results of microscopic simulations $\eta$ are reported in Section~\ref{sec:simulations}. Finally, conclusions are drawn in the last Section. In this Section we discuss also the differences between the present and the original model of Refs.~\cite{bagnoli2013, bagnoli2015}.

\begin{figure}[t]
\begin{center}
\includegraphics[width=\columnwidth]{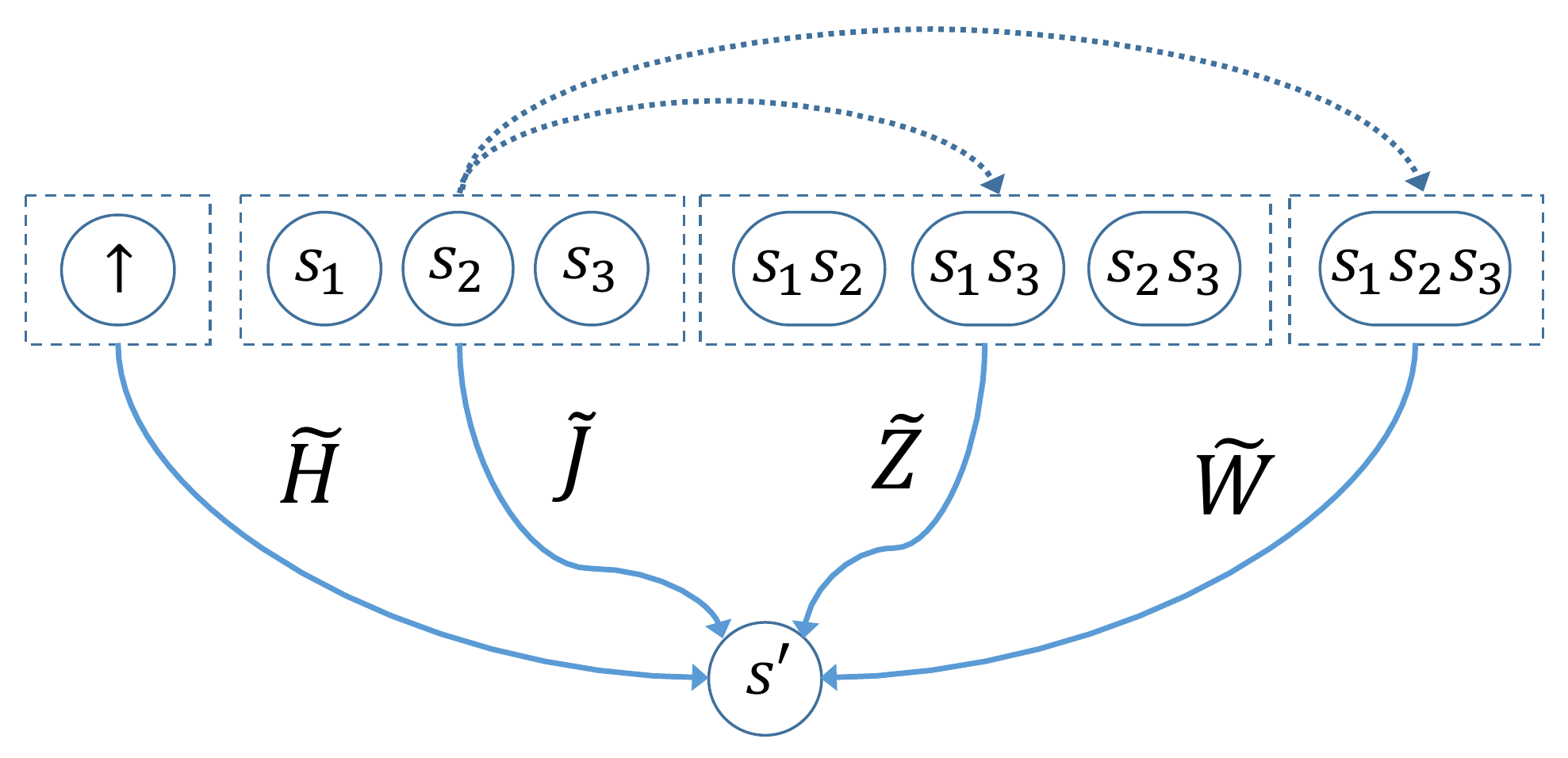}
\end{center}
\caption{\label{plaquette} (Color online.) A $K=3$ spin neighborhood, with the interaction terms corresponding to the external field $\tilde H$,  the
two-spin  $\tilde J$, three-spin $\tilde Z$, and four-spin $\tilde W$ interaction constants.}
\end{figure}


\section{Parallel nonlinear Ising model}\label{sec:parallelising}
We consider a system with $N$ sites, each one in a state $s_i\in
\{-1,1\}$, $i=1,\dots,N$. The state of the system is
$@s=(s_1,\dots,s_{N})$.  The topology of the
system is defined by the adjacency matrix $a$ with $a_{ij}=1$ if site
$j$ belongs to site $i$'s neighborhood and is zero otherwise. The connectivity $k_i$, the local field $\tilde{h}_i$, and the rescaled local
field $h_i$ at site $i$ are
\[
 k_i=\sum_j a_{ij},\qquad
  \tilde{h}_i=\sum_j a_{ij}s_j, \qquad
  h_i=\dfrac{\tilde{h}_i}{k_i},
\]
with $h_i\in [-1,1]$. In this paper we shall use a uniform connectivity $k_i=K\,\,\forall\, i$. The magnetization  $m$ is defined as
\[
 m=\dfrac{1}{N}\sum_is_i,
\]
with $m\in [-1,1]$.

In the following we consider multi-spin (plaquette) interactions. We moreover consider only completely asymmetric interactions~\cite{Suzuki, Sherrington}, arranged to give a preferred direction that corresponds to time in the standard cellular automata language~\cite{bagnoli2013, SS-bif}.  

Considering up to 4-spin interactions, the Hamiltonian is
\begin{align}\label{Hamiltonian}
    \mathcal{H} (@s) =& -\sum_i s_i'\Bigl(\tilde H+ \tilde J\sum_j a_{ij} s_j +%
        \tilde Z \sum_{jk} a_{ij} a_{ik}s_j s_k + \nonumber\\
		&\tilde W \sum_{jkl} a_{ij} a_{ik} a_{il} s_j s_k s_l\Bigr),
\end{align}
where $\tilde H$ is the external field, and
  $\tilde J$, $\tilde Z$,  $\tilde W$ the two-spin, three-spin and four-spin interaction constants respectively as shown in Fig.~\ref{plaquette}.

It is possible to recast the interaction constants in terms of the local field $h_i$. 
The terms containing $s'_i$ at ``time'' $t+1$ are, 
\[ 
\begin{split}
&\text{2-spin: }\; s_i'\sum_j a_{ij} s_j = s_i' \tilde h_i,\\  
&\text{3-spin: }\; s_i'\sum_{jk} a_{ij} a_{jk}s_j s_k=s_i'Q^{(2)}_i,\\
&\text{4-spin: }\; s_i'\sum_{jkl} a_{ij} a_{ik} a_{il} s_j s_k s_l  = s_i'Q^{(3)}_i.\\
\end{split}
\]
These expressions define $ Q^{(2)}$ and $ Q^{(3)}$. Since
\[
\begin{split}
&\tilde{h_i}^2 = K + 2Q_i^{(2)},\\
&\tilde{h_i}^3 =  (3K-2)\tilde h_i + 6 Q_i^{(3)},\\
\end{split}
\]
the Hamiltonian can be written as
\begin{equation}
\mathcal{H} (@s) = -\sum_i s'_i
		(H + J h_i  + Zh_i^2 + W h_i^3),
\end{equation}
where the correspondences among coupling constants are
\[
\begin{split}
H &= \tilde H - \frac{1}{2} K\tilde Z,\\
J  &= K\left(\tilde J - \frac{3K-2}{6}\tilde W\right),\\
Z &= \frac{1}{2} K^2\tilde Z, \\
W &= \frac{1}{6}K^3\tilde W.\\
\end{split}
\]
In the following, we shall consider pair ($J$) and four-spin ($W$) terms, \textit{i.e.}, $H=Z=0$, in agreement with previous investigations~\cite{bagnoli2013}.

The coupling term $J$ modulates the ``linear'' effects of neighbors, so $J>0$ gives a conformist (ferromagnetic) behavior and $J<0$ a contrarian (antiferromagnetic) one. The term $W$ modulates the nonlinear effects of the crowd.  In this way one can model the Asch effect by inserting $J<0$ (contrarian attitude) and $W>0$ (social norms).  

The time evolution of the spins is given by the parallel application of 
the transition probabilities $\tau(s_i'|h_i)$ that gives the probability that the spin at site $i$ and time $t+1$ takes value  $s_i'$  given  the local field $h_i$ at time $t$, see Fig.~\ref{plaquette}. The local transition probability is defined by a heat bath  probability 
\begin{align}\label{eq:ntau}
  \tau(s'_i| h_i) &= \dfrac{1}{1+\exp(-2 s'_i (J h_i + W h_i^3))}\nonumber\\
  &=\frac{1}{2}\left[1+s'_i\tanh (J h_i + W h_i^3)\right].
\end{align}

\begin{figure}[t]
  \centering
  \begin{tabular}{cc}
    (a) & (b)\\
   \includegraphics[width=0.45\columnwidth]{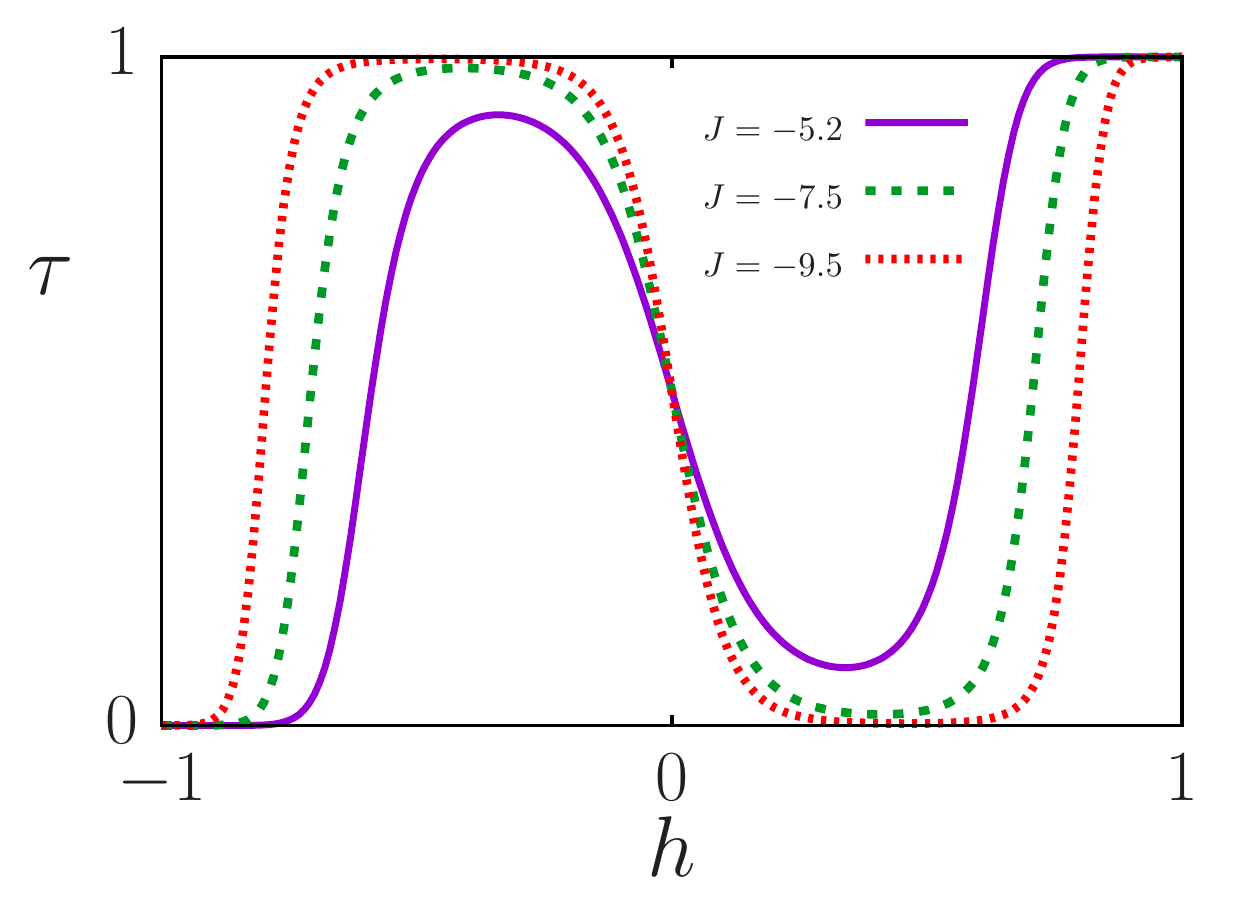} &
   \includegraphics[width=0.45\columnwidth]{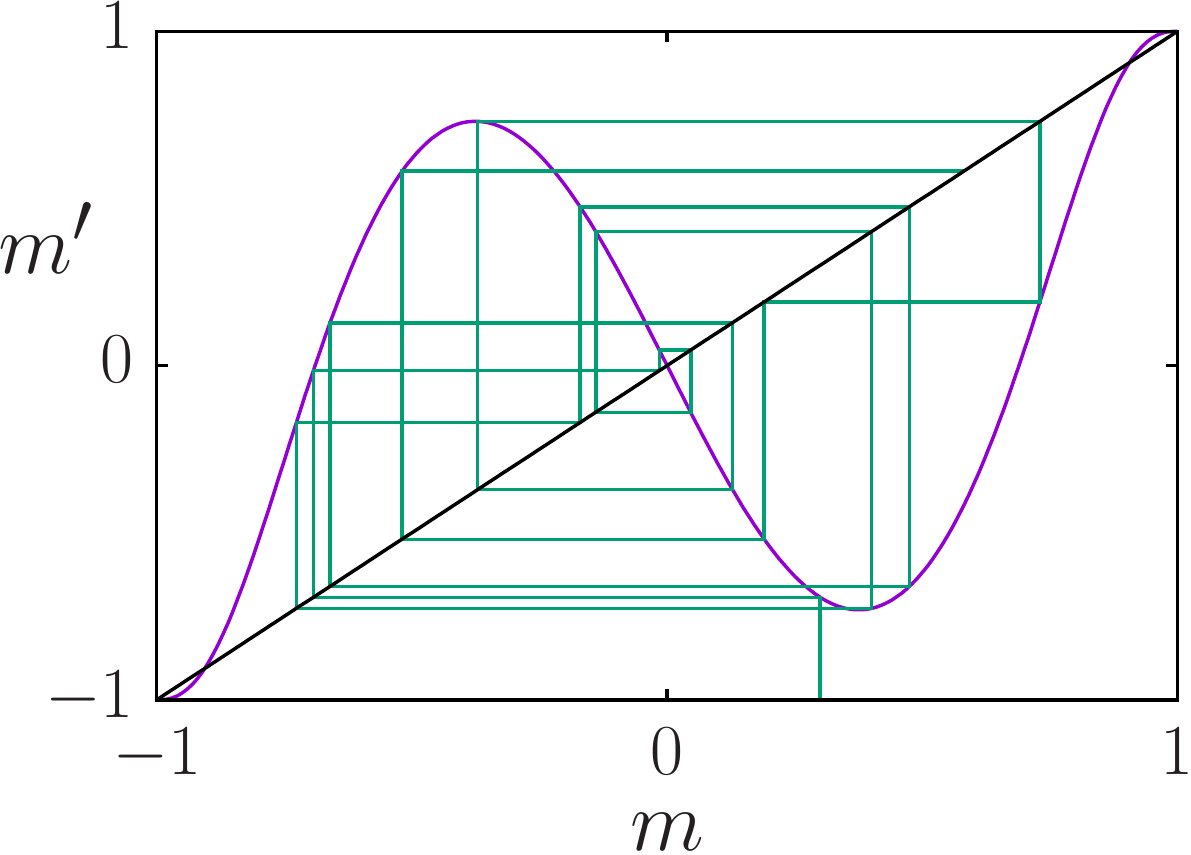}
  \end{tabular}
 \caption{\label{fig:tau} (Color online). (a) The transition
   probability $\tau=\tau(1|h)$, Eq.~\eqref{eq:ntau} as a function of the local
   field $h$ for three values of the coupling constant $J$. (b) Graph of the magnetization  $m'$ at time $t+1$ as a function of the magnetization $m$ at time $t$,  Eq.~\eqref{eq:mf}, with some iterates for
   $J=-7.5$. }
\end{figure}

The parallel version of the linear ($W=0$) Ising model does not show many differences with respect to the standard serial one~\cite{Derrida}. The observables that depend only on single-site properties take the same values in parallel or sequential dynamics~\cite{NewmannDerrida}, although differences arise for two-site correlations~\cite{SS-meta}. 
In general the resulting dynamics is no more reversible with respect to the Gibbs measure induced by any Hamiltonian~\cite{Cirillo}. 

In the following, unless otherwise specified, we always use $W=15$ and $K=20$. 


\begin{figure}[t]  
  \centering{
  \begin{tabular}{cc}
    (a) & (b) \\
  \includegraphics[width=.45\columnwidth]{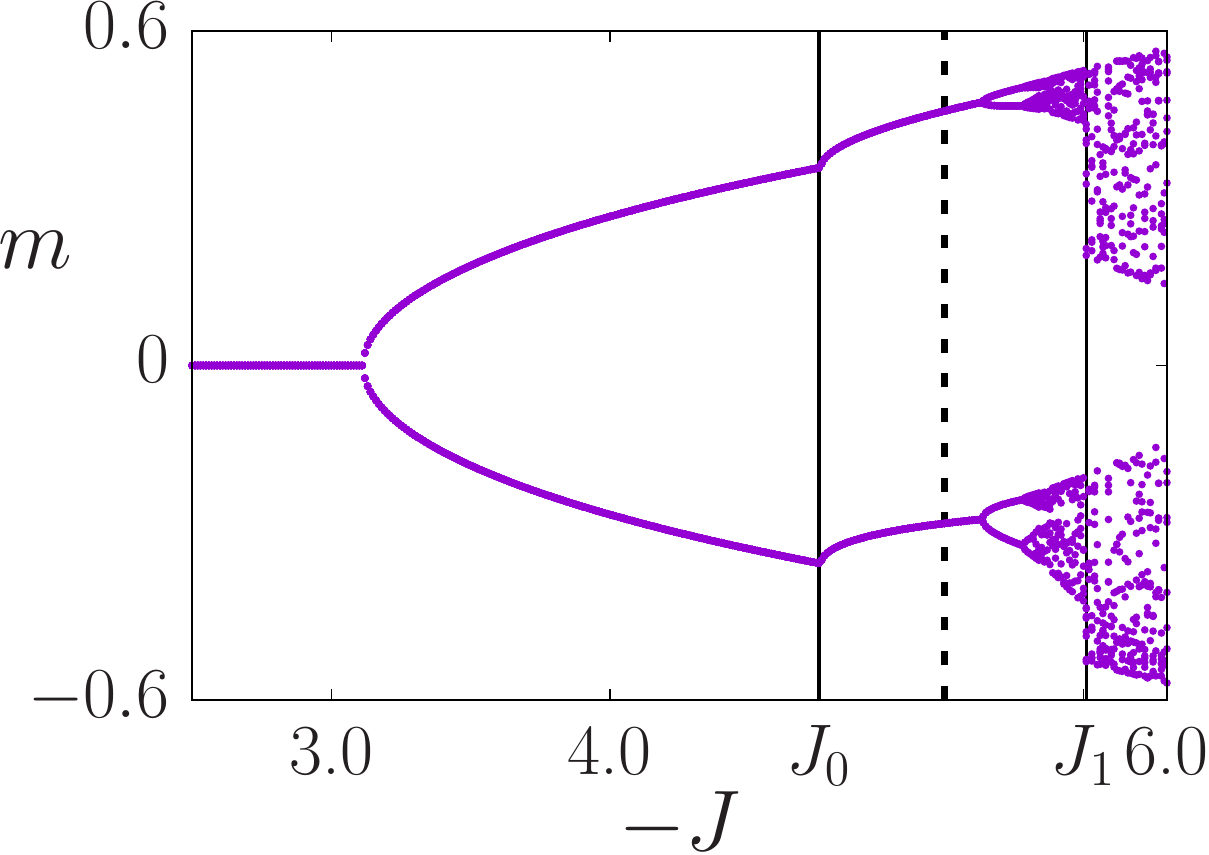} &
  \includegraphics[width=.45\columnwidth]{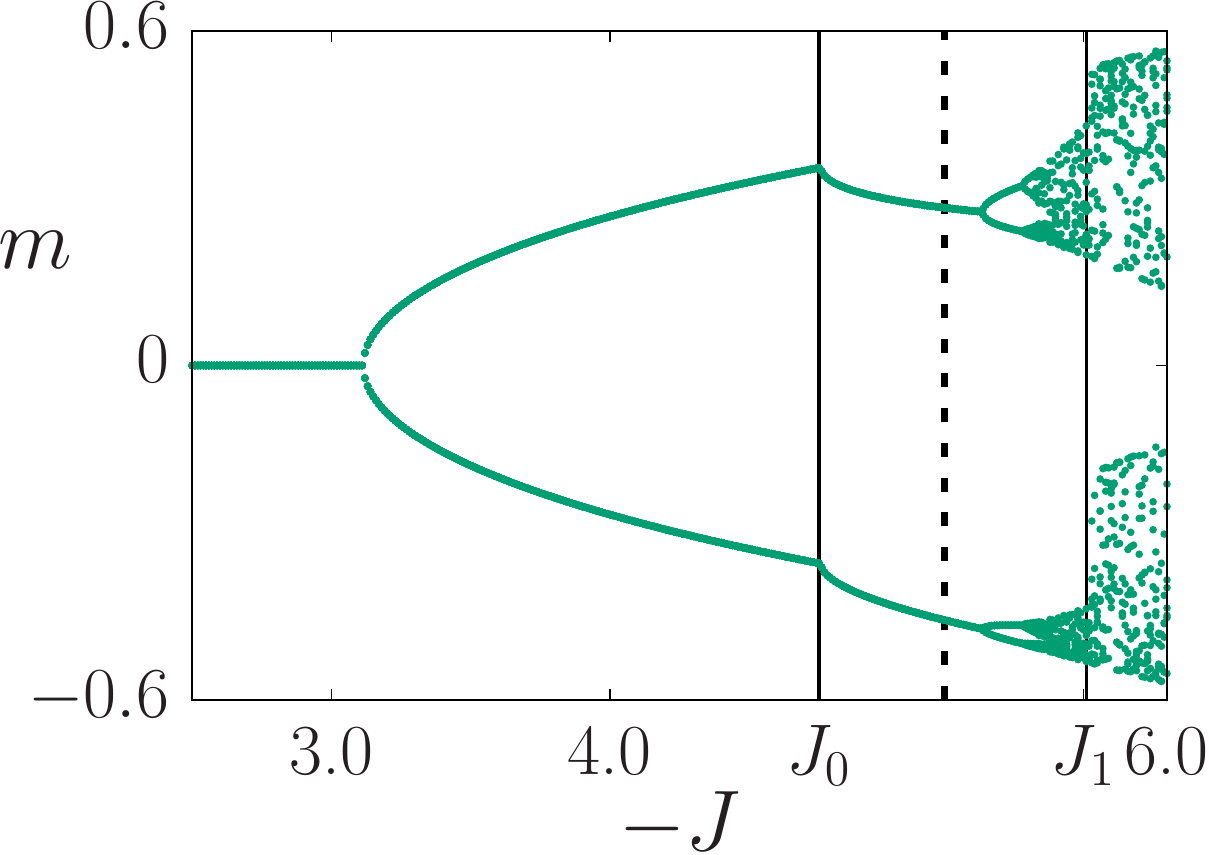} 
  \end{tabular}\\[2mm]
  (c) \\
  \includegraphics[width=.8\columnwidth]{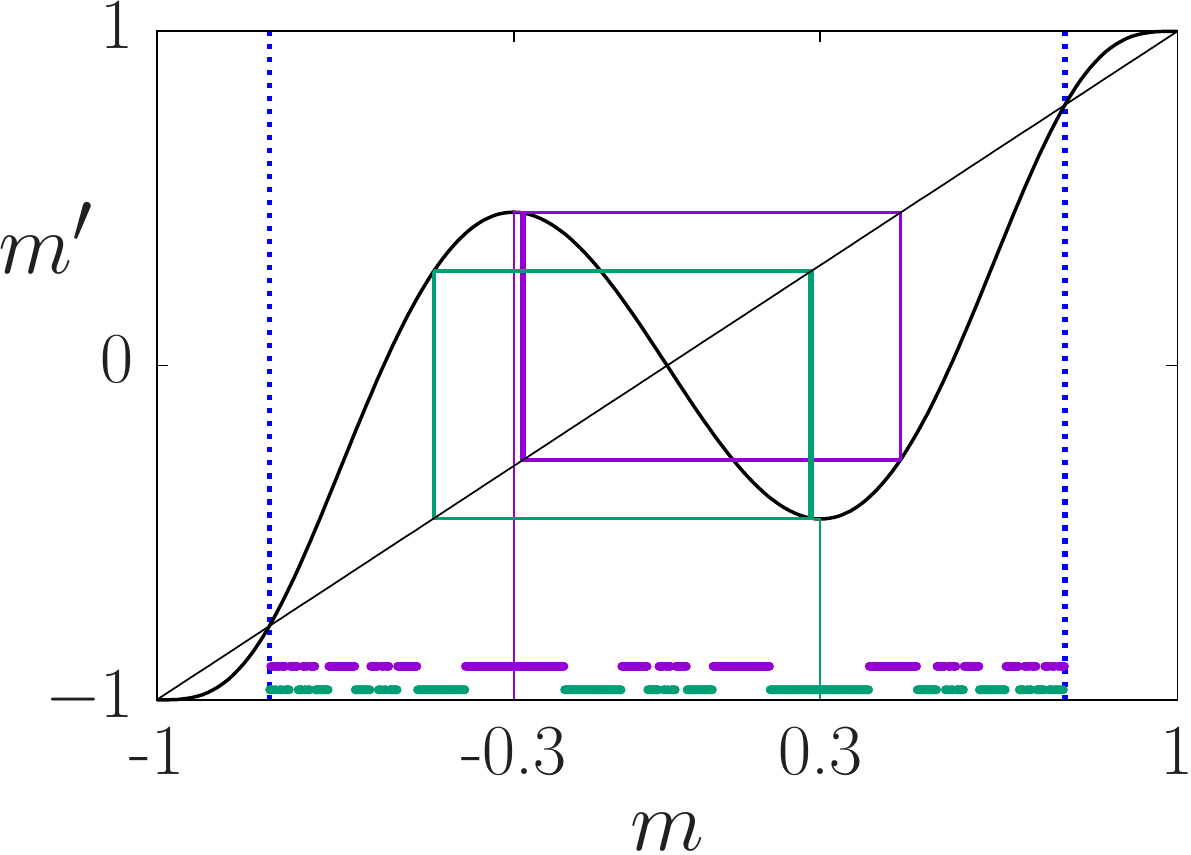}
  }
  \caption{\label{fig:mf-bif-c0p60} (Color online). 
  (a) Mean-field
    bifurcation diagrams of the magnetization $m$ for initial
    magnetization $m_0=-0.3$ in (a) and $m_0=0.3$ in (b), as functions
    of the linear coupling $J$. In (a) and (b), the vertical dotted lines (in black) are drawn at
    $J=-5.2$ and the vertical continuous lines (in black) at $J_0=-4.75$ and $J_1=-5.71$. Between these values,
    the bifurcation diagram depends on the initial magnetization $m_0$. In (c) the return map of Eq~\eqref{eq:mf} for $J=-5.2$ 
    corresponding to the  dotted line in the previous figures.  The
    vertical  lines  mark the basins of attraction of the two period-2 orbits. The dots
    in the bottom part (in green) mark the basin including $m_0=0.3$ corresponding to the leftmost orbit (also in green) and figure (b), 
    while the dots  above them (in magenta) mark the basin including $m_0=-0.3$, corresponding to the rightmost orbit
    (also in magenta) and figure (a).}
\end{figure}

\section{Mean-field approximation}\label{sec:meanfield}

The mean-field approximation for the magnetization $m$ of the fully
parallel case with fixed connectivity $K$, follows from the Markov
equation assuming no spatial correlations. Then
\begin{align}\label{eq:mf}
 m'=&f(m) = \frac{1}{2^{K}}\sum_{k=0}^K\binom{K}{k}(1+m)^k(1-m)^{K-k}\nonumber\\
 &\times\tanh\left[J\left(\frac{2k}{K}-1\right)+W\left(\frac{2k}{K}-1\right)^3\right],
\end{align}
with $m=m(t)$ and $m'=m(t+1)$. We show in Fig.~\ref{fig:tau} (b) the
graph of $m'$ together with some iterates of the map.

The mean-field magnetization  exhibits chaos that can be characterized by
the Lyapunov exponent $\lambda$~\cite{ott02}.  However, on spatially extended networks
$m$ changes stochastically 
and cannot be characterized in the same way.

In order to
compare microscopic and mean-field models within the same framework we
use the Boltzmann's entropy $\eta$~\cite{Boltzmann,bagnoli2013} of the
magnetization $m$. The interval $[-1,1]$ is partitioned in $L$
disjoint intervals $I_i$ of equal size and the probability $q_i$ of
$I_i$ is the fraction of visits to $I_i$ after $T$ time steps with
$T\gg 1$. 

Once
these probabilities are known, $\eta$ is defined by
\begin{equation}
 \label{eq:s}
 {\eta}=- \dfrac{1}{\log L}\sum_{i=1}^Lq_i\log q_i,
\end{equation}
so that $0\leq \eta\leq 1$, the lower bound corresponding to
a fixed point, the upper one to the uniform distribution
$q_i=1/L$.

 For a periodic orbit of period $2^p$ and $L=2^b$,
$\eta=p/b$.  For low-dimensional dynamical systems, like the
mean-field equation, the mid-value threshold $\eta=0.5$ effectively
separates the contracting dynamics (cycles) from the chaotic ones. For
spatially-extended systems, there is always a stochastic noise that
increases the value of the entropy in the ``fixed-point'' part of the
parameter space. This base-level value is related to the size of the
sample, and   slowly vanishes
for large samples. 

In order to use finite-size samples, we set the
onset of the phase in the stochastic systems corresponding to the
chaotic phase in the deterministic ones to the mid-value of the range
of $\eta$. Taking the limits $T\to\infty$, $L\to\infty$ leads
to the Kolmogorov-Sinai entropy~\cite{kolmogorov58,kolmogorov59,sinai59,ott02}. 

Before presenting the different  scenarios, let us illustrate the type of bifurcations 
that are present. In Figs.~\ref{fig:mf-bif-c0p60} (a) and (b) we show parts of the bifurcation diagram of the map of
Eq.~\eqref{eq:mf} as a function of $J$ staring with different values of the initial magnetization $m_0$.

Referring to the values in the Figure, at $J=J_0$ there is a pitchfork bifurcation, i.e., a separation of basins, 
that reunite at $J=J_1$, which is another pitchfork bifurcation, in the reverse direction. Intermixed, there are period-doubling
bifurcations.  There are other 
pitchfork bifurcations for different intervals of $J$.

In Fig.~\ref{fig:mf-bif-c0p60} (c) we show the return map of the mean-field map
for $J=-5.2$. We can see that there are four basins of attraction. For small values of the initial magnetization 
$m_0$, the orbit is attracted to $m=-1$ and for large $m_0$ to
$m=1$. The regions where this occurs are marked by the vertical
dotted lines in the Figure. For other values of $m_0$, $m$ ends in one
of two period-two orbits. This figure shows, in the lower part, the
two basins of attraction that are symmetric in the sense that if $m_0$
belongs to one basin of attraction, $-m_0$ belongs to the other one.

In what follows we present the bifurcation diagrams of the  mean-field map Eq.~\eqref{eq:mf}
by varying the coupling constants $J$ and $W$. Unless otherwise noticed, we computed the Lyapunov exponent 
$\lambda$ by averaging over 10,000 time steps after a transient of another $10,000$ steps. The entropy $\eta$ was computed 
using 256 boxes and $25,000$ time steps.

\begin{figure}[t]    
  \centering
 \includegraphics[width=0.9\columnwidth]{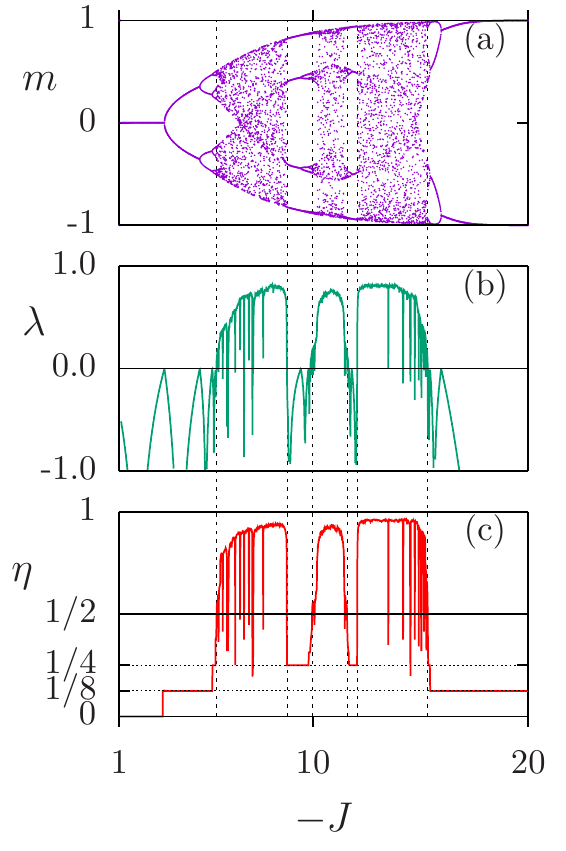} 
  \caption{\label{fig:mf-J} (Color online.) (a) Bifurcation diagram of
    the mean-field map of the magnetization $m$, Eq.~\eqref{eq:mf}, as
    a function of the linear coupling $J$. In (b)
    the corresponding Lyapunov exponent $\lambda$ and in (c) the entropy
    $\eta$.  
    The vertical dotted lines are drawn at the estimated values of $J$
   for which  $\lambda=0$. The horizontal dotted lines in (c)
    correspond to period 2, $\eta=1/8$, and period 4,
    $\eta=1/4$, orbits. For every value of $J$, two initial values of the magnetization were
    used, $m_0=-0.3$ and $m_0=0.3$.}
\end{figure}
\begin{figure}[t]  
  \centering
 \includegraphics[width=0.9\columnwidth]{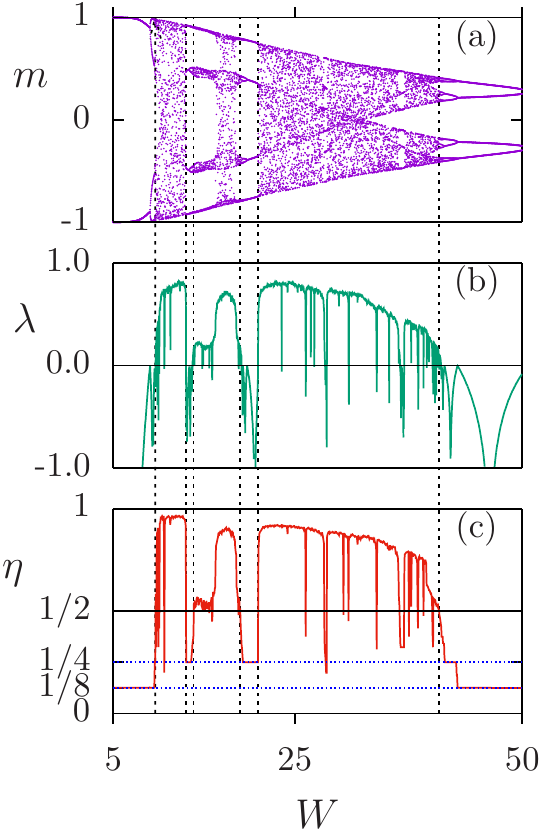}
  \caption{\label{fig:mf-W} (Color online.) (a) Bifurcation diagram of
    the mean-field map of the magnetization $m$, Eq.~\eqref{eq:mf}, as
    a function of the coupling constant $W$ with  $J=-10$. In (b)
   The corresponding Lyapunov exponent $\lambda$ and in (c) the entropy
    $\eta$.  
    The vertical dotted lines are drawn at the estimated values of $W$ 
    for which $\lambda=0$. The horizontal dotted lines in (c)
     correspond to period 2, $\eta=1/8$, and period 4,
    $\eta=1/4$, orbits. For every value of $W$, two initial values of the magnetization were
    used, $m_0=-0.3$ and $m_0=0.3$.}
\end{figure}

In Fig.~\ref{fig:mf-J} (a) we show the bifurcation diagram of the
magnetization $m$, Eq.~\eqref{eq:mf}, as a function of $J$ with $W$
and $K$ fixed. The diagram exhibits a period doubling cascade towards
chaos with periodic windows and pitchfork bifurcations. 

The
bifurcation diagram of $m$ as a function of $W$ with $J$ and $K$ fixed is shown in
Fig.~\ref{fig:mf-W} (a).  In this case, there is an inverse period
doubling cascade to chaos with pitchfork bifurcations.

The next row of figures,
Figs.~\ref{fig:mf-J} (b) and Fig.~\ref{fig:mf-W} (b), show the corresponding  Lyapunov
exponent $\lambda$, and the last one, Figs.~\ref{fig:mf-J} (c) and Fig.~\ref{fig:mf-W} 
(c), the entropy $\eta$. The dotted vertical lines are drawn at some
of the values of $J$ or $W$ where $\lambda$ passes from a negative to
a positive value or vice versa.  These values coincide to jumps of
$\eta$ from values smaller to $1/2$ to larger ones or vice
versa and mark the appearance of chaos or periodic
windows in the bifurcation diagrams.

\begin{figure}[t]
  \begin{center}
    \includegraphics[width=0.9\columnwidth]{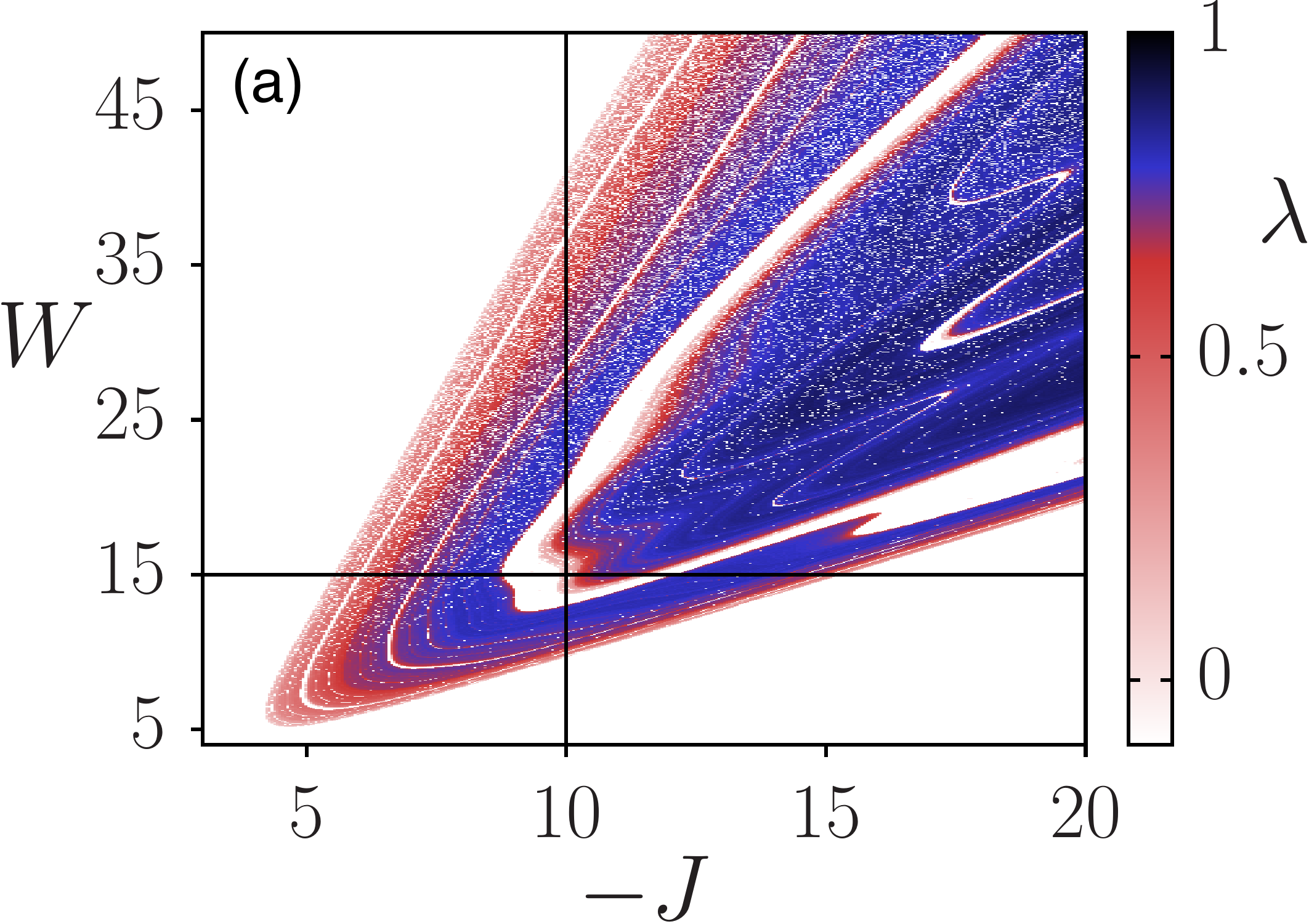}\\[3mm]
    \includegraphics[width=0.9\columnwidth]{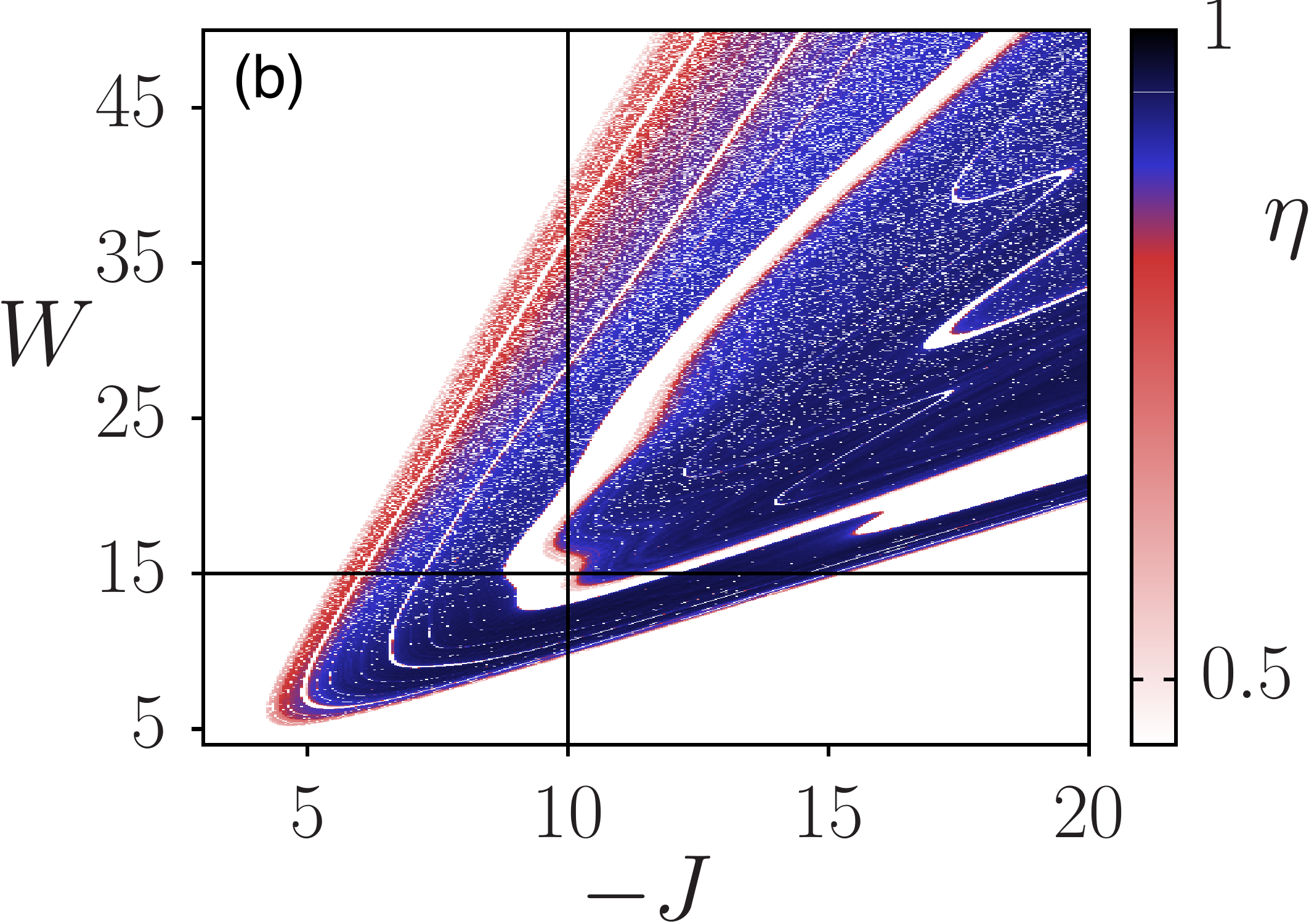}
  \end{center}  
  \caption{\label{fig:mfpdWJ} (Color online) (a) Mean-field phase
    diagram of the Lyapunov exponent $\lambda$ showing the values of
    $(-J,W)$ where $\lambda>0$.  (b) The mean-field phase diagram
    of the entropy showing the values of $(-J,W)$ where
    $\eta>1/2$. }
\end{figure}

Therefore, $\eta$ can be used as a
measure of chaos.  To stress this, we show in Figs.~\ref{fig:mfpdWJ}
the mean-field phase diagrams of $\lambda$ (top) and $\eta$ (bottom).
These diagrams are similar. The horizontal lines at
$W=15$ correspond to Figs.~\ref{fig:mf-J} (b) and (c) and the vertical
ones at $J=-10$ to Figs.~\ref{fig:mf-W} (b) and (c) respectively.  We find a
similar behavior of the mean-field map as $K$ varies with fixed $J$
and $W$. The three quantities $J$, $W$ and $K$ are related by scaling
relations, as shown in the Appendix.

The dotted horizontal lines in Fig.~\ref{fig:mf-J} (c) and  Fig.~\ref{fig:mf-W} (c)
correspond to period 2, $\eta=1/8$, and period 4, $\eta=1/4$
orbits. Looking at the bifurcation diagram in Fig.~\ref{fig:mf-J} (a),
for small $-J$, the map has a fixed point and as $-J$ grows there is a
first bifurcation to a period 2 orbit and another one to what looks
like a period 4 orbit, but $\eta=1/8$ instead of $\eta=1/4$ for period
4 orbits. What appears like a bifurcation to period four orbits is
actually a pitchfork bifurcation to two period-two orbits that depend
on the initial magnetization $m_0$ as mentioned before.  
There are other pitchfork bifurcations for other values of $J$ with
$W$ fixed and also for values of $W$ with $J$ fixed.


\section{Small-World stochastic bifurcations}\label{sec:simulations}

In the Watts-Strogatz small-world model~\cite{WattsStrogatz}, starting
with a network where every site has $K$ nearest neighbors,
at any site $i$, with probability $p$, known as the long-range
connection probability, each one of its $K$ neighbors is replaced by
a random one.  Then the spin at each site is
updated according to Eq.~\eqref{eq:ntau}. As $p$ grows, coherent oscillations
of a majority of spins begin to appear so that the magnetization $m$
shows noisy periodic or irregular oscillations. The noise is the
manifestation of the stochasticity of the updating rule.  Similar
patterns can be seen in Ref.~\cite{bagnoli2005}, where the effect of
the size of neighborhood is studied.

As shown in the following, by changing several parameters, we can
obtain stochastic bifurcation diagrams similar to the mean-field ones.
The following microscopic simulations were carried using lattices of $N=10,000$ sites, 
with a transient of $10,000$ time steps. The entropy $\eta$ was computed with 256 boxes and $25,600$ time steps.

\begin{figure}
 \begin{center}
 \includegraphics[width=\columnwidth]{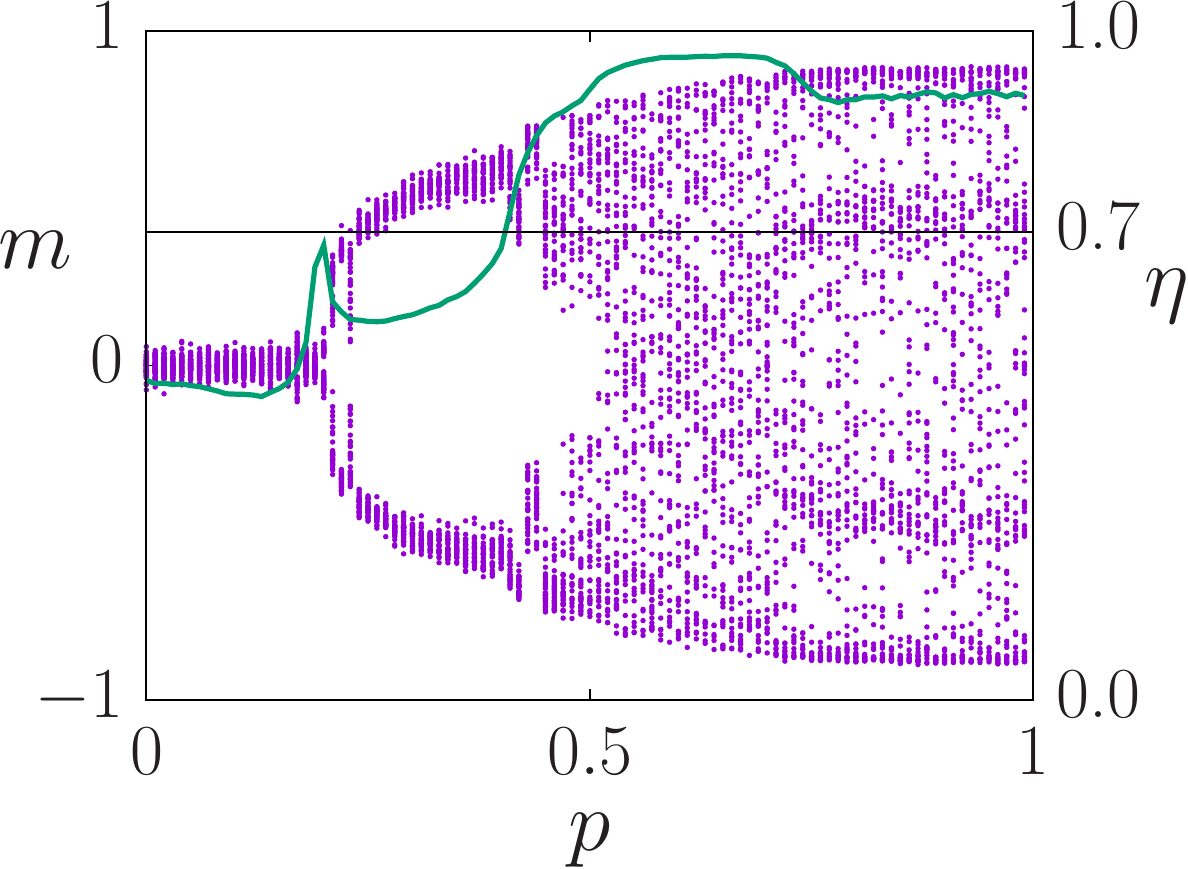}
 \end{center}
 \caption{\label{bifp} (Color online) Small-world stochastic
   bifurcation diagram of the magnetization $m$, dots (in magenta),
   and the entropy $\eta$, continuous curve (in green), as functions
   of the long-range connection probability $p$ with $J=-10$. 
   The ``jump'' of $m$ for $p\simeq 0.45$ corresponds to a pitchfork
    bifurcation (dependence on the initial magnetization).}
\end{figure}

In Fig.~\ref{bifp} we show the bifurcation map and the entropy $\eta$
as functions of $p$. There is always some disorder, even for small
values of $p$ where $m\sim 0$, and as $p$ grows we find bifurcations and
more disorder. Indeed, the entropy $\eta$ is a good measure of this behavior, small values of
$\eta$ corresponding to noisy ``periodic orbits'' while larger
ones to disorder (``chaos''). 

In the mean-field case, we found $\eta=1/2$ to be a
good threshold to separate order from chaos.  For the stochastic
dynamics on small-world networks we choose as the threshold the
approximate value of the entropy at the first bifurcation as shown in
the figures. For values smaller than this
threshold there are noisy periodic orbits.  
The bifurcation diagram of the
figure is reminiscent of the mean-field one, Fig.~\ref{fig:mf-J}.

Notice that pitchfork bifurcations (dependence on the initial magnetization) are present also 
in the microscopic simulations, as shown in Fig.~\ref{bifp}

\begin{figure}
  \begin{center}
    \begin{tabular}{cc}
      $p=0.3$ & $p=0.5$ \\
      \includegraphics[width=0.45\columnwidth]{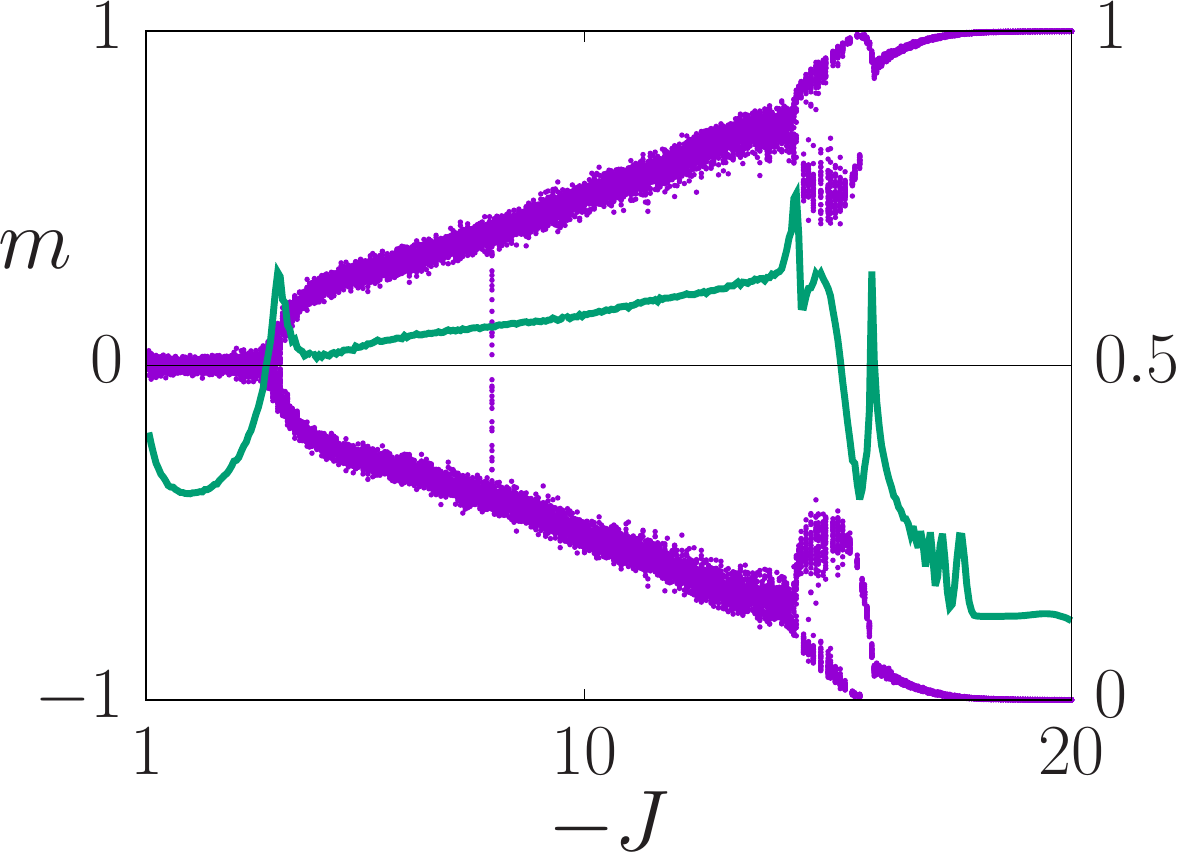} &
      \includegraphics[width=0.45\columnwidth]{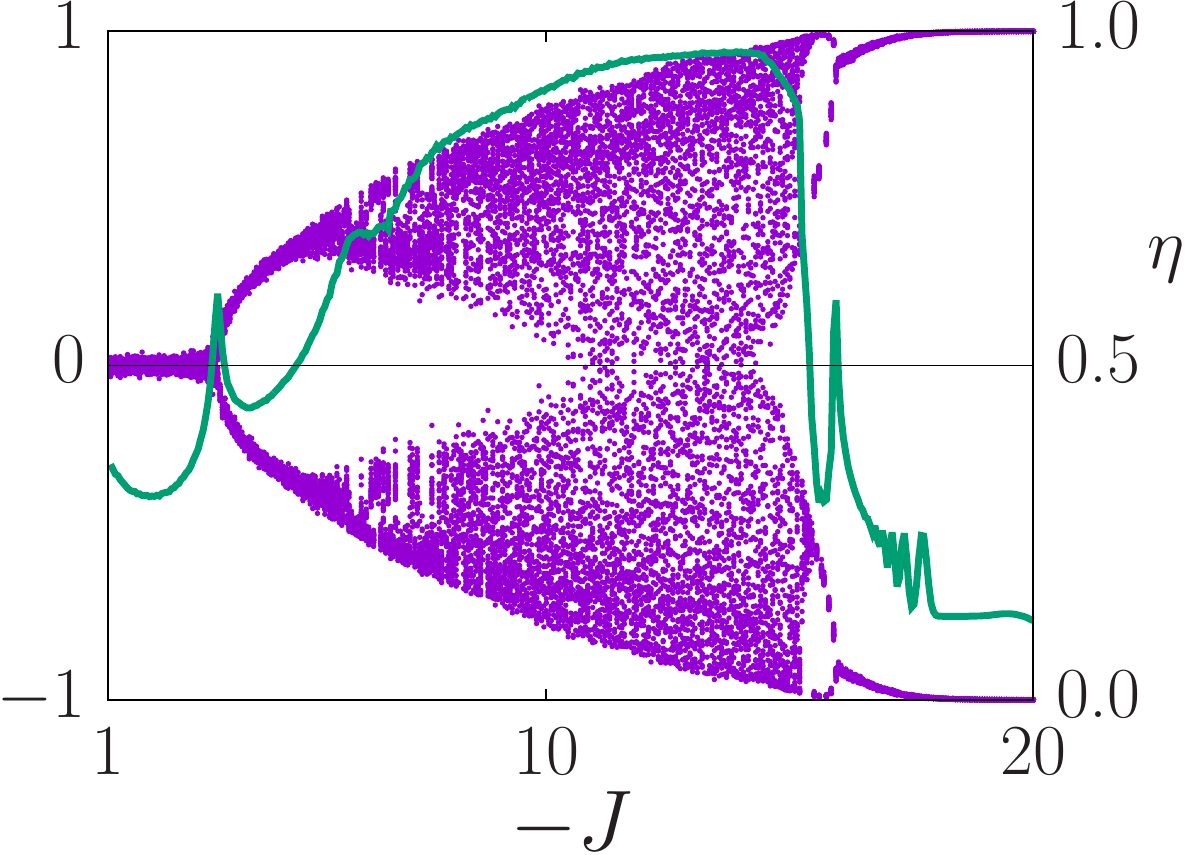} \\
   $p=0.8$ & $p=1.0$ \\                 
     \includegraphics[width=0.45\columnwidth]{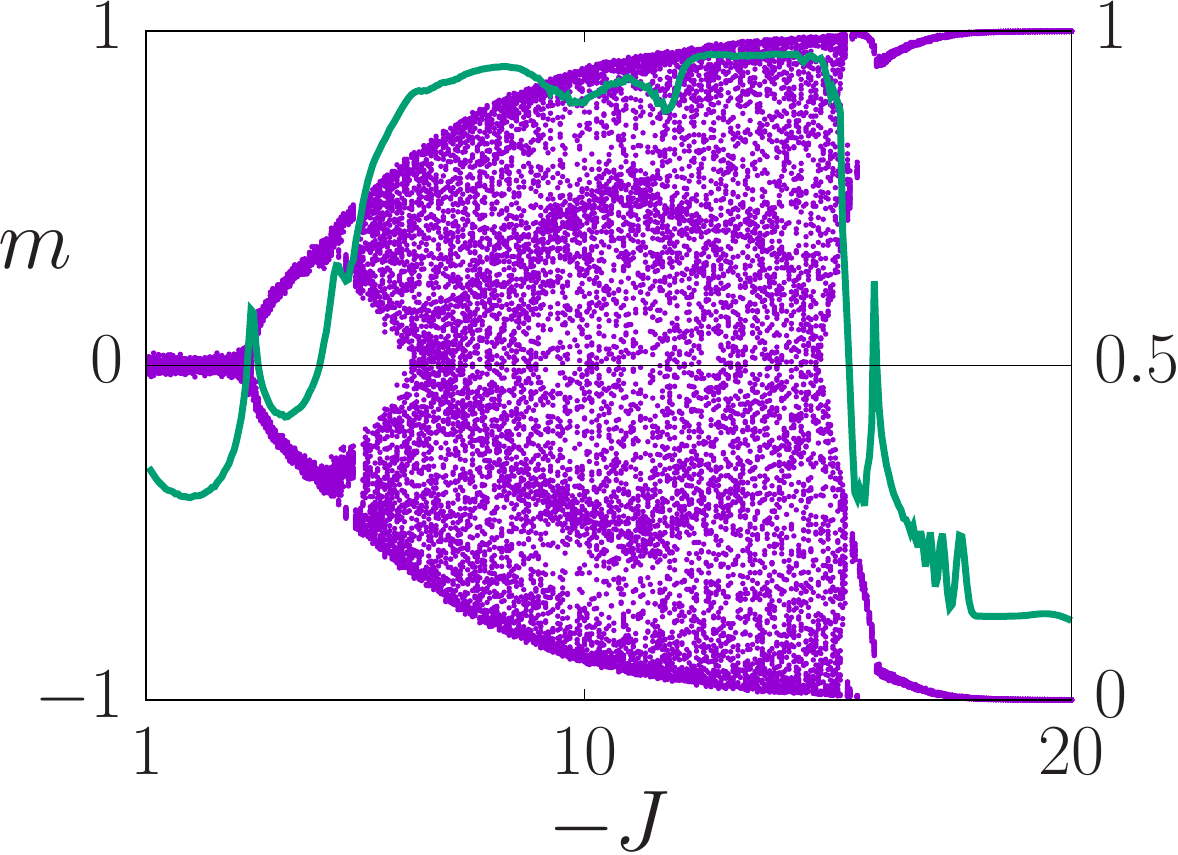} &    
     \includegraphics[width=0.45\columnwidth]{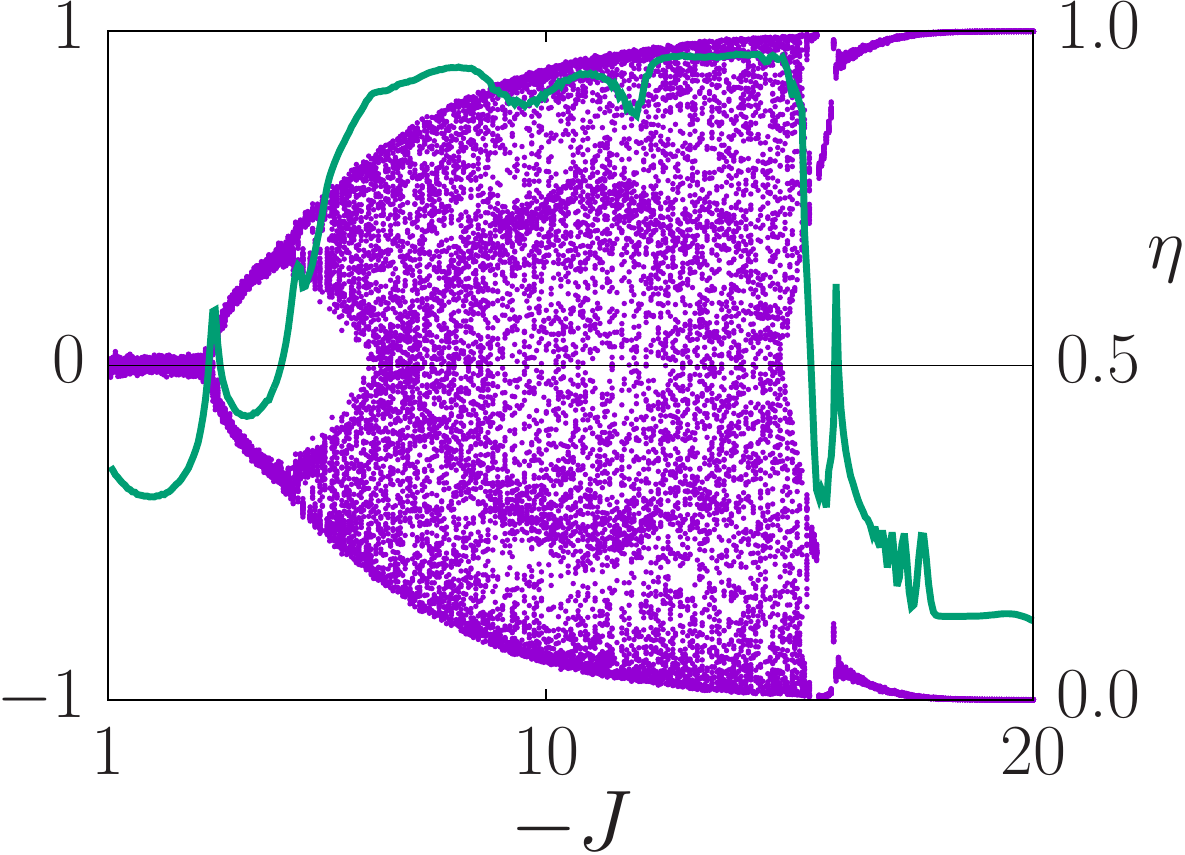}
    \end{tabular}
   \end{center} 
 \caption{\label{fig:sw-J} (Color online) Small-world bifurcation diagram
   of the magnetization $m$, dots (in magenta) and the corresponding
   entropy $\eta$, continuous line (in green), as functions of of the
   linear coupling constant $J$  and different
   values of the long-range connection probability $p$.}
\end{figure}

In Fig.~\ref{fig:sw-J} we show the bifurcation diagrams and entropy
of the magnetization $m$ for the small-world networks obtained for different
values of $p$. As before, the entropy is a good indicator of
disorder.


Clearly, by setting the rewiring probability $p$ large enough, one can also recover 
the mean-field bifurcation diagrams as function of $J$, $K$ and $W$, with a good correspondence of the critical values of parameters.


\subsection{Partial asynchronism (dilution)}

The dilution $d$  is the fraction of
sites chosen at random that are not updated at every time step.
We define the diluted rule as 
\begin{equation}\label{p}
s_{i}(t+1) = \begin{cases}
               1 & \text{with probability  $ (1-d)   \tau(1|h_i)$,}\\
               -1 & \text{with probability  $(1-d)   \left[1-\tau(1|h_i)\right]$,}\\
               s_{i}(t) & \text{otherwise, i.e., with probability $d$,}
               \end{cases}
 \end{equation}
so that for $d=0$ one has the standard parallel updating rule. 
One time step is defined when on the average every site of the lattice
is updated once. For a system with $N$ sites, the smallest value of
the dilution is $d=1/N$ and then $t_d=1/d$ updates are needed to
complete one time step. If $d=1/2$, $t_d=2$, etc.

\begin{figure}
 \begin{center}
   \includegraphics[width=\columnwidth]{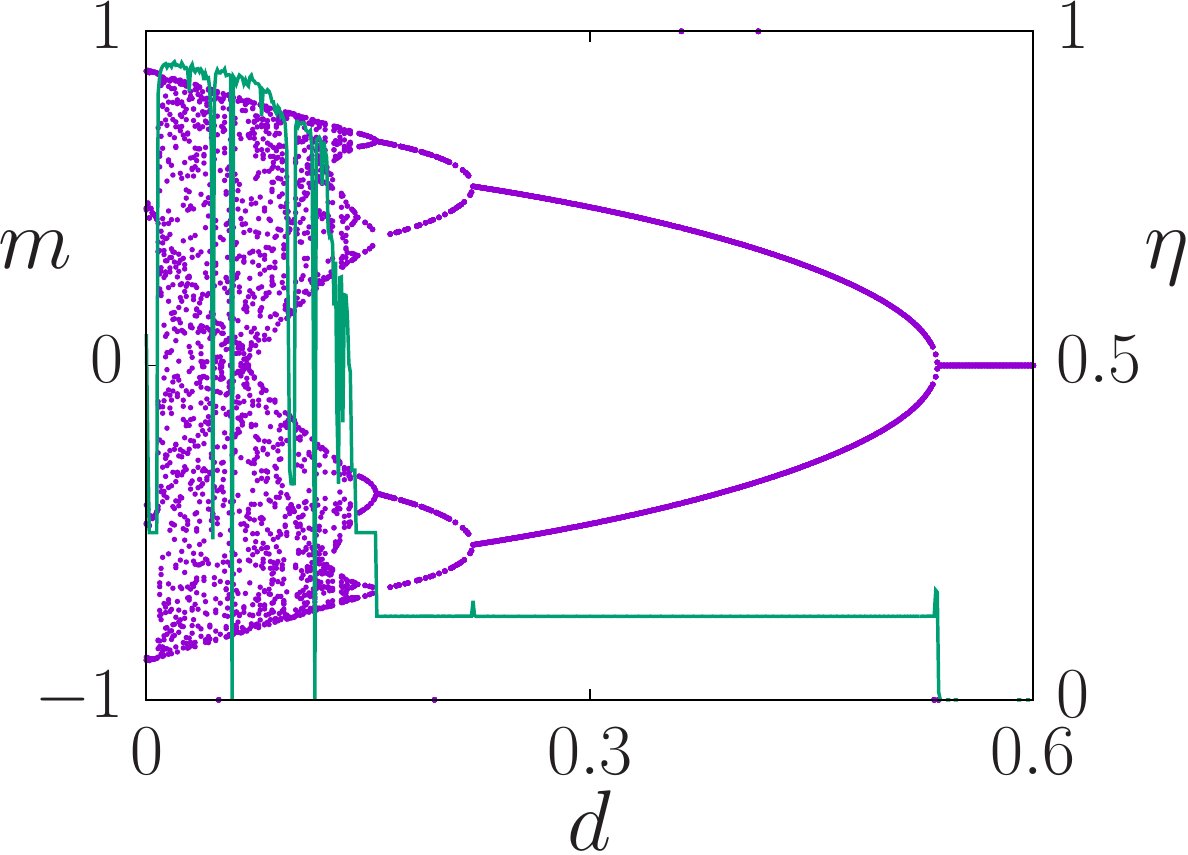}
 \end{center}
 \caption{\label{fig:mfpddJ} (Color online) Mean-field
   bifurcation diagram of the magnetization $m$  (dots in magenta, two initial conditions), and the
   entropy $\eta$  (continuous curve in green), as 
   functions of the dilution probability $d$ with 
   $J=-10$. }
\end{figure}

The mean-field equation corresponding to dilution is 
\[
	m(t+1) = (1-d)m(t) + d f(m(t));
\]
where $f$ is the the map of Eq.~\eqref{eq:mf}. The
mean-field phase diagram is reported in Fig.~\ref{fig:mfpddJ}. Notice
that the border at $d=0$ corresponds to the horizontal line in
Fig.~\ref{fig:mfpdWJ}.

\begin{figure}[t]
  \begin{tabular}{cc}
   $p=0.1$ & $p=0.2$ \\                 
    \includegraphics[width=0.45\columnwidth]{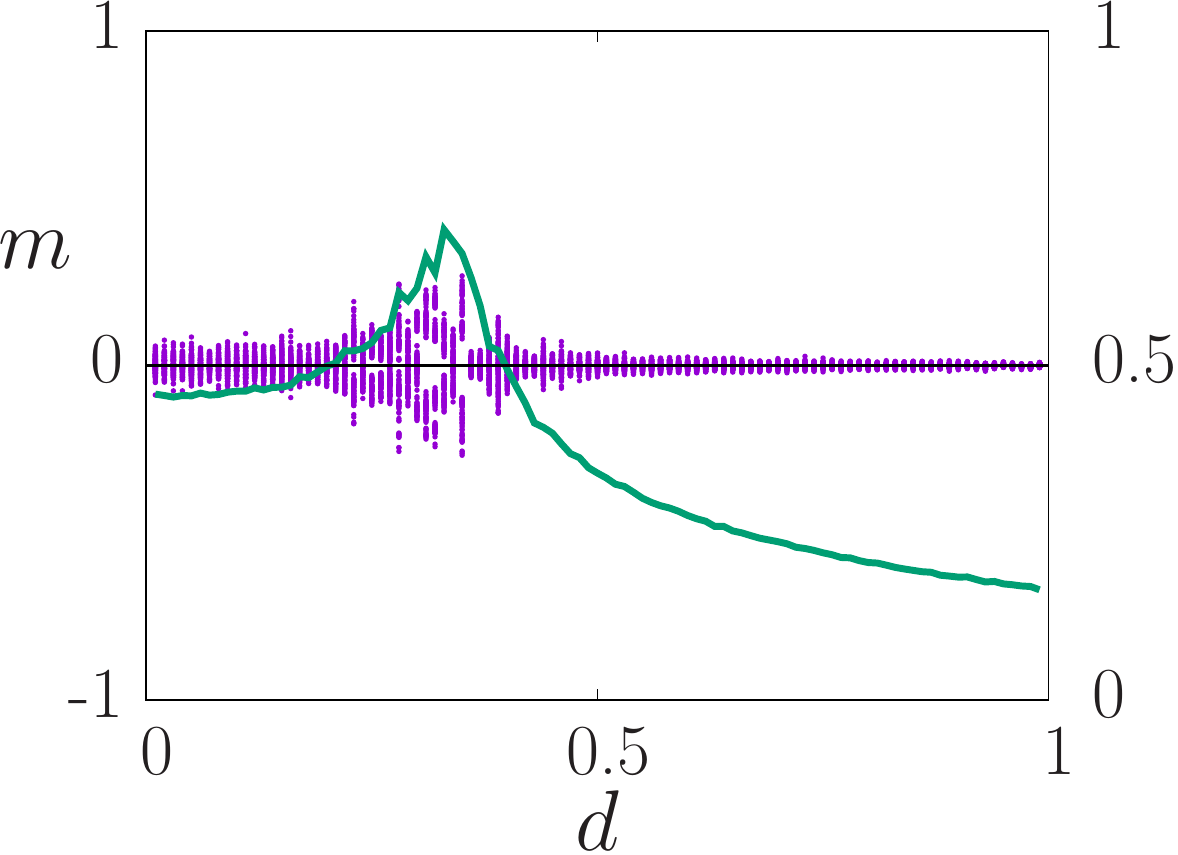} &
    \includegraphics[width=0.45\columnwidth]{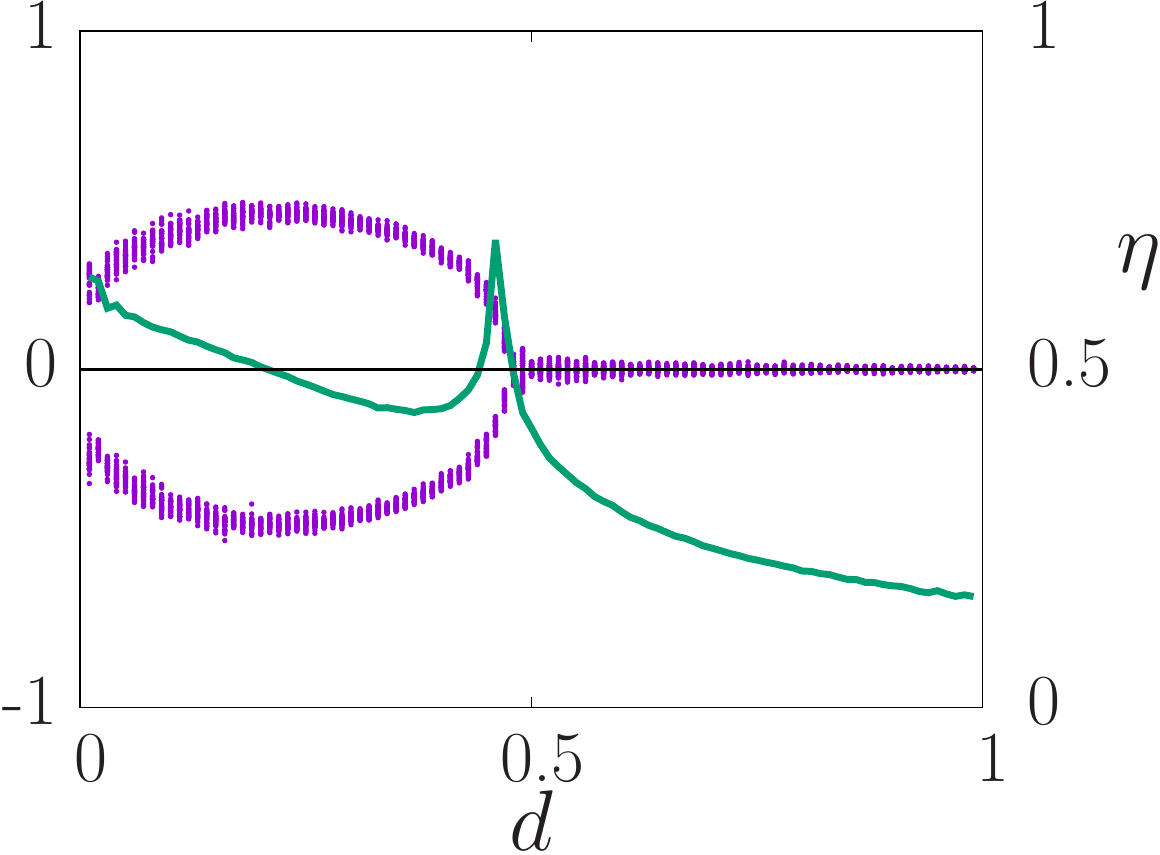}\\
   $p=0.5$ & $p=1.0$ \\                 
    \includegraphics[width=0.45\columnwidth]{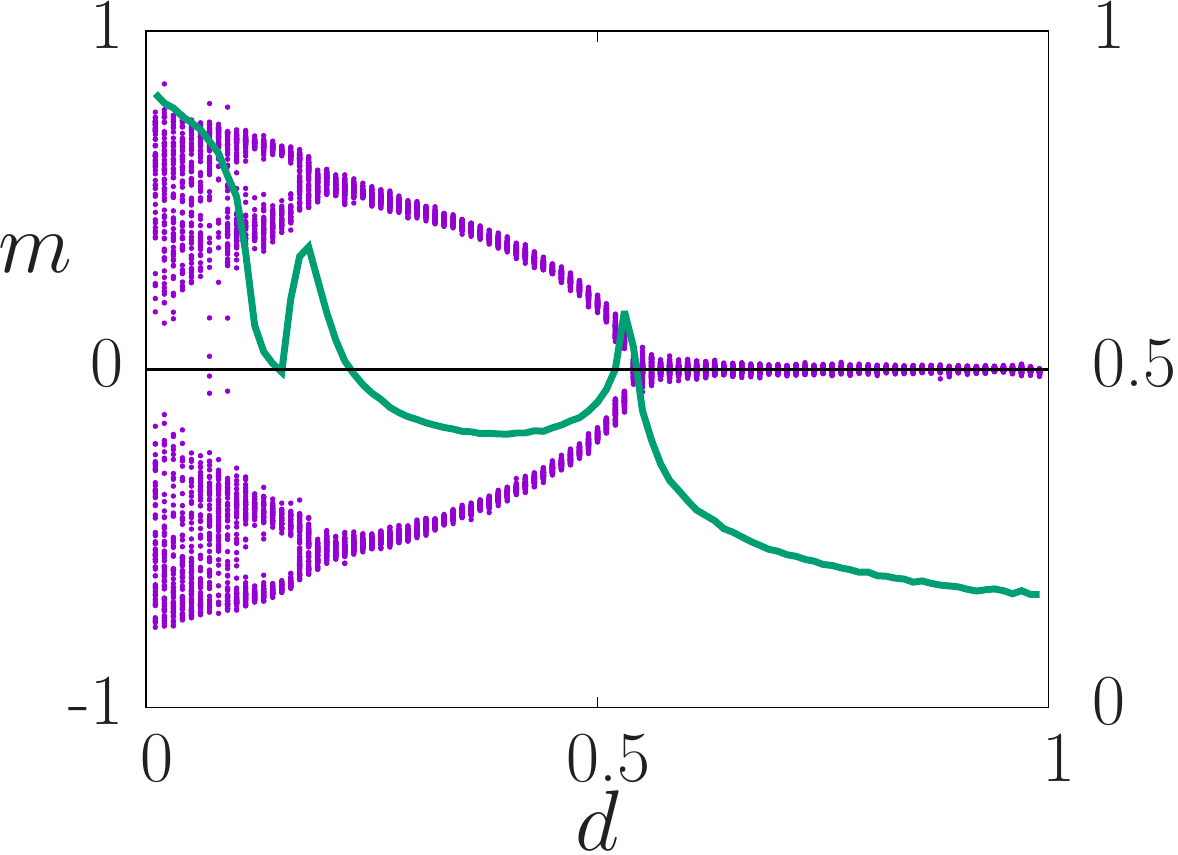} &
    \includegraphics[width=0.45\columnwidth]{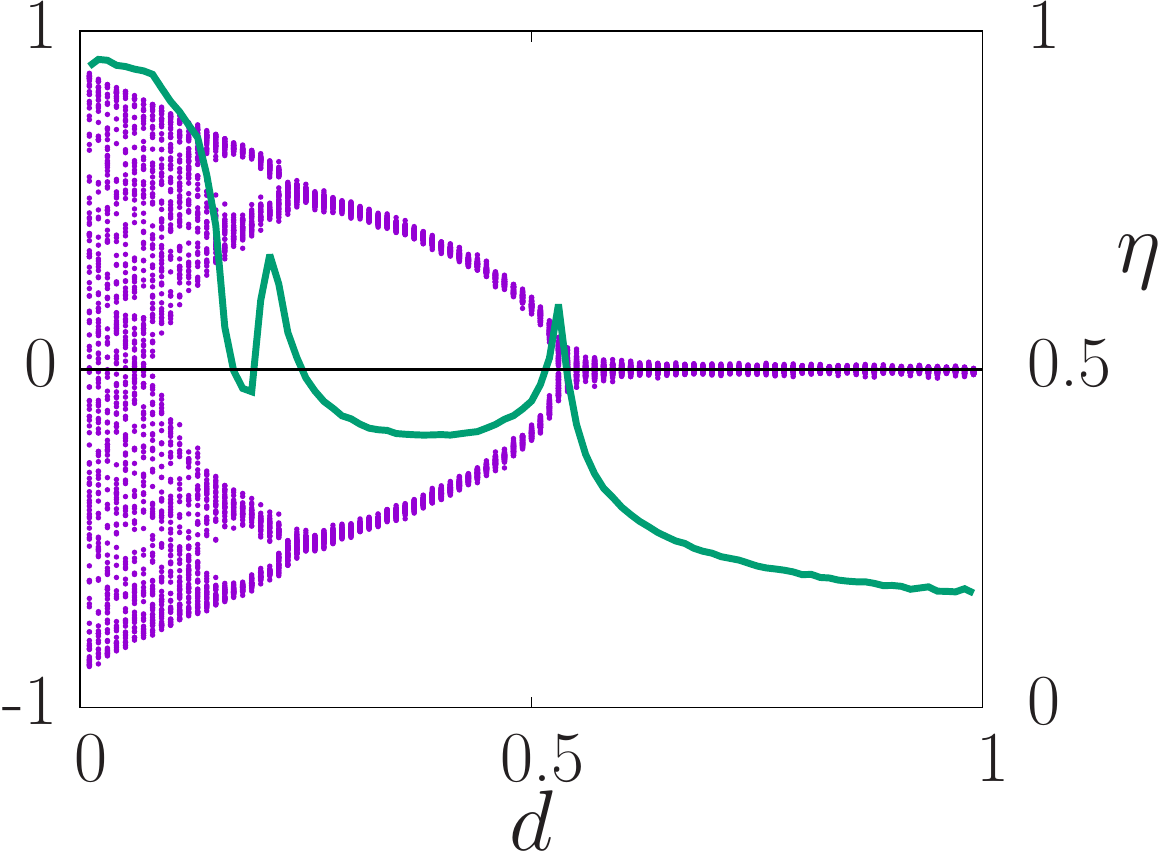} \\
  \end{tabular}
  \caption{\label{bifd} (Color online). Bifurcation diagrams of the
    magnetization $m$ on small-world networks, dots (in magenta), and
    the entropy $\eta$, continuous curve (in green), as functions of
    the dilution $d$ with $J=-10$, and different
    values of the long-range connection probability $p$.}
\end{figure}

The bifurcation diagrams and the entropy $\eta$ of the magnetization as functions of
of the dilution $d$ are shown in Fig.~\ref{bifd} for different value of the long-range
connection probability $p$. It is interesting to note the ``bubbling''
transition: the oscillations are favored, for intermediate values of
the rewiring $p$, by a non-complete parallelism. 

As shown in the figure, for values of $p$ larger than $0.1$, the dilution is able to trigger
bifurcations also in the spatial model.  In contrast with  the linear Ising
model~\cite{Cirillo}, where even a small amount of asynchronism is
able to destroy the ``effective'' antiferromagnetic coupling, here the
behavior is smooth with respect to
dilution. See Ref.~\cite{SS-meta} for a study about metastable effects
in the linear model.

\subsection{Heterogeneity}

\begin{figure}
  \begin{tabular}{cc}
   $p=0.05$ & $p=0.2$ \\
    \includegraphics[width=0.45\columnwidth]{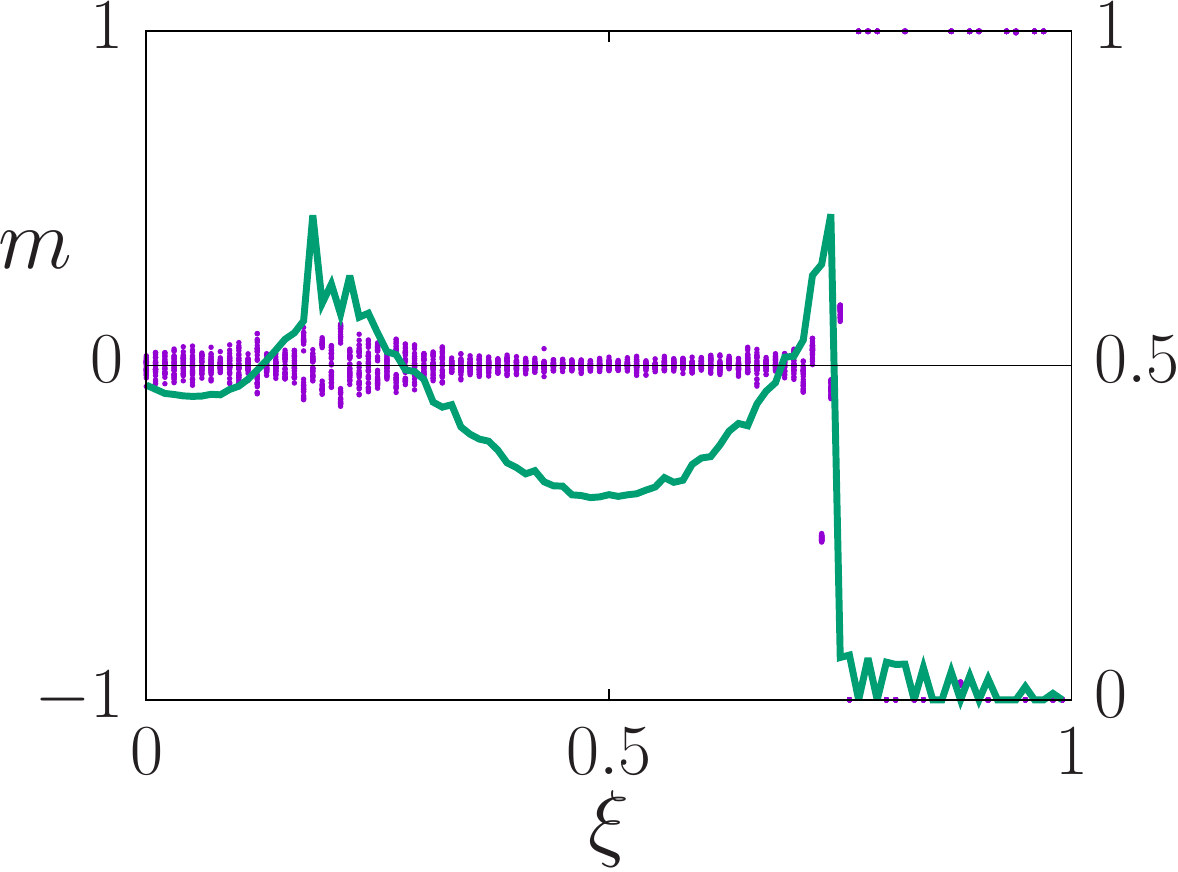} &
    \includegraphics[width=0.45\columnwidth]{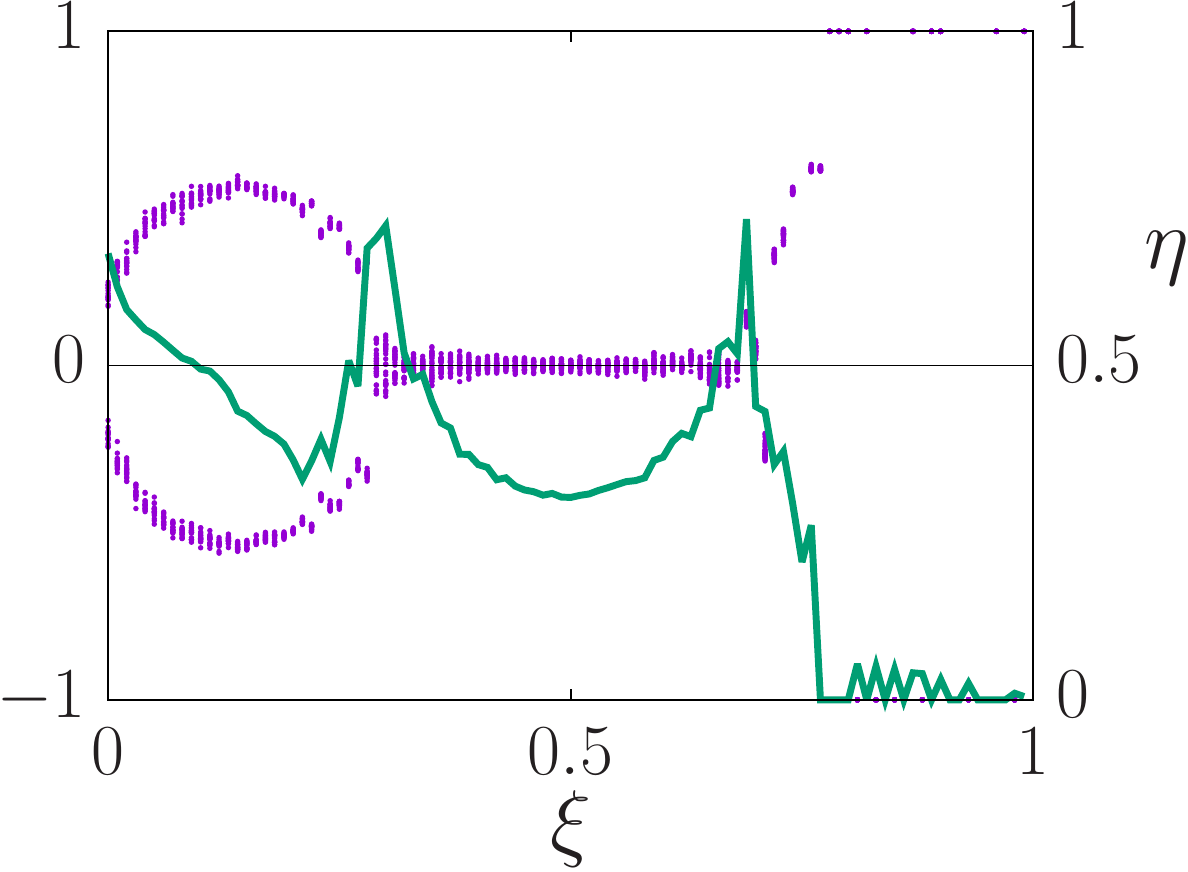}\\
    $p=0.8$ & $p=1.0$\\                    
    \includegraphics[width=0.45\columnwidth]{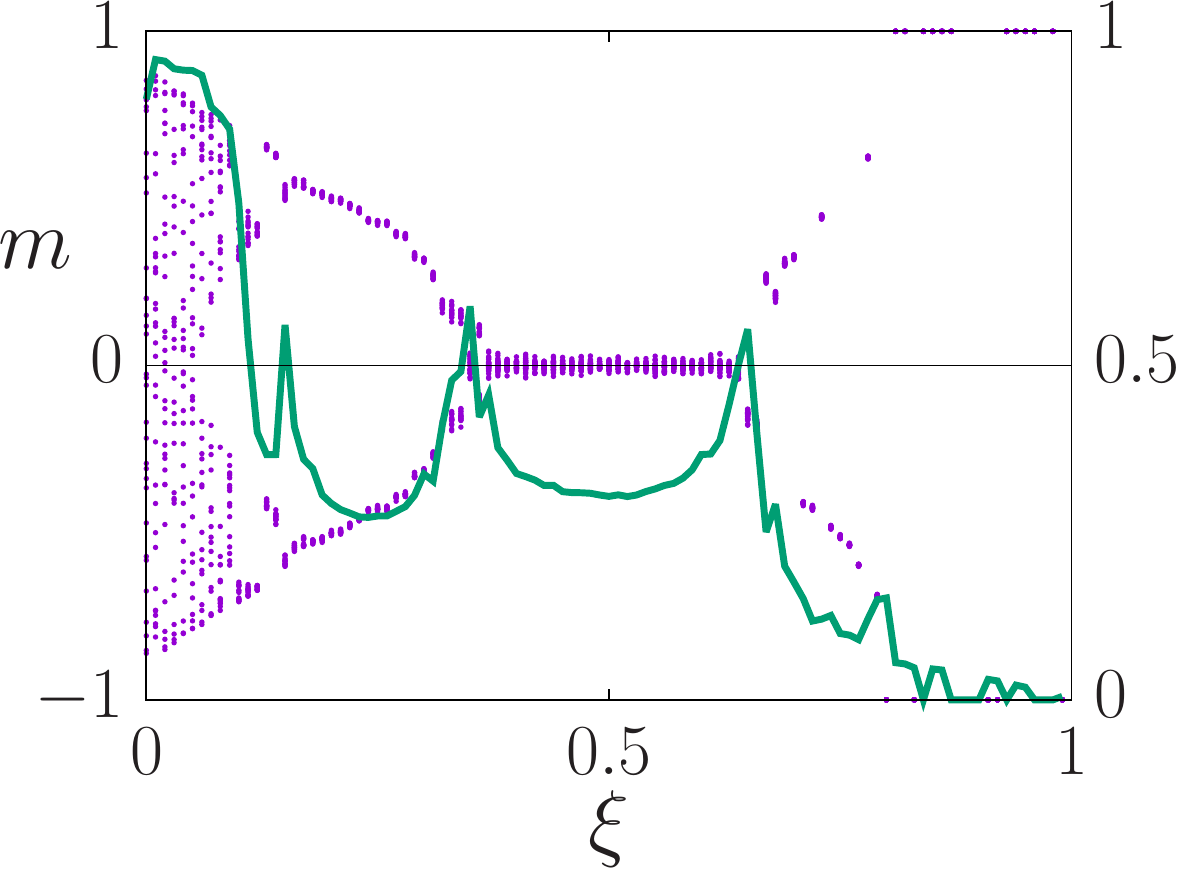} &
    \includegraphics[width=0.45\columnwidth]{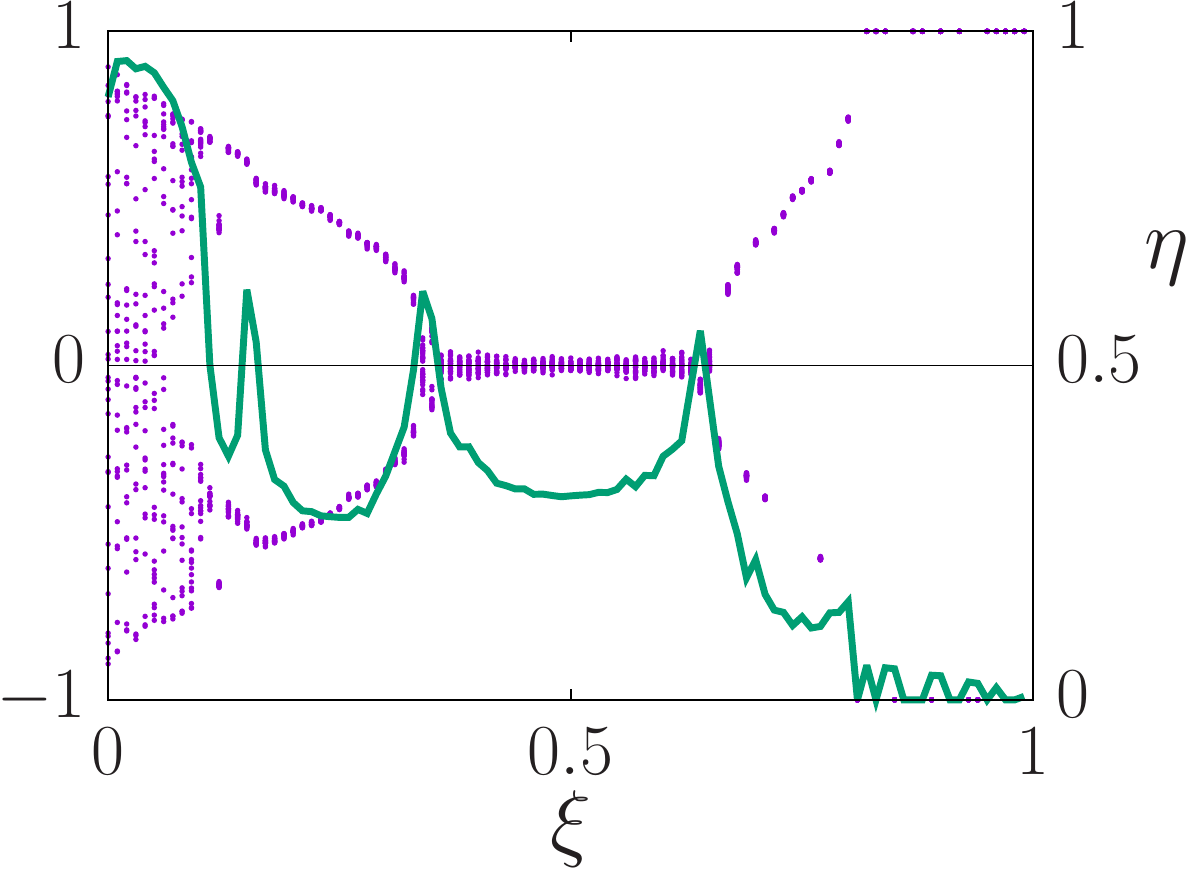} 
  \end{tabular}
  \caption{\label{fig:mix} (Color online.) Small-world ferro-anti ferro
    bifurcation diagram (left axis, dots in magenta) and entropy $\eta$ (right axis, continuous curve in green) of
    the magnetization $m$ as a function of $\xi$ for different values
    of the long range probability $p$ and $N=10,000$,
    $|J|=10$,. for $P=0.05$ there is no threshold for $\eta$, for $p=0.2$ it is
    $\eta=0.7$, for $p=0.8$ and $p=1.0$ it is $\eta=0.6$.}
\end{figure}

In order to measure the effects of heterogeneity, we 
let a fraction $\xi$ of spins interact
ferromagnetically ($J>0$) with their $K$ neighbors and a fraction
$1-\xi$ interact antiferromagnetically ($J<0$). We show in
Fig.~\ref{fig:mix} the bifurcation diagrams of the magnetization $m$
together with the entropy $\eta$ as functions of $\xi$ for different
values of $p$. Again, the entropy is a good measure of disorder.
One can see a ``bubbling'' effect very similar to
what observed by changing the dilution. In other words, oscillations,
which are a product of antiferromagnetism and parallelism, are
actually favoured by a small percentage of asynchronism and/or of
ferromagnetic nodes, for a partial long-range rewiring of links.


\section{Conclusions}\label{sec:conclusions}

We investigated the phase transitions of a nonlinear, parallel version
of the Ising model, characterized by a linear coupling $J<0$ and a
nonlinear one $W>0$.  The mean-field approximation shows chaotic
oscillations, by changing the couplings $J$ and $W$ or the
connectivity $K$.  We  showed in the Appendix that there is a
scaling relation among these parameters.

The nonlinear Ising model was  studied on small-world networks, where $p$ is the probability of long-range rewiring of links.
Here, entropy of the magnetization becomes a measure of disorder which is adequate once a
threshold between the presence and absence of noisy periodic orbits is
established. The noisy periodic and disordered behavior of $m$ imply a
certain degree of synchronization of the spins, induced by long-range
couplings. 

We have shown also that similar bifurcations may be induced in the
randomly connected model by changing the parameters $J$, the dilution factor $d$ and the heterogeneity $\xi$,
by mixing ferromagnetic and antiferromagnetic interactions.

In particular, we observed that  a small percentage of asynchronism or of ferromagnetic nodes favours the first  
period-doubling bifurcation, which appears first for intermediate values of $d$ and $\xi$. 

This model is a generalization of an opinion formation model presented in Refs.~\cite{bagnoli2013} 
and \cite{bagnoli2015}. In contrast with those investigations, we developed here the 
whole model within the framework of the parallel Ising model, with transition probabilities that are continuous and smooth, derived from Hamiltonian couplings. We found the mean-field bifurcation and phase diagrams as functions of 
$J$ and $W$, and discussed the dynamics on small-world networks as functions of the long-range rewiring probability $p$ and the heterogeneity $\xi$ The results are qualitatively similar 
to those found before, showing a certain degree of universality regardless of the details of the model. 

We obtained also new results, 
such as the mapping among the parameters (only possible within this continuous approach) 
and the stochastic bifurcation phase diagram 
as function of the asynchronism (or dilution) $d$ of the updating rule.

The different diagrams show a striking similarity, implying that it should be possible to map one bifurcation onto the other, as we did 
withing the mean-field approach among $J$, $W$ and $K$.

The present model aims at incorporating the Asch effect~\cite{Asch} in mean-field and microscopic simulations, 
{\em i.e}, the influence of social pressure and its dominance over the ``linear'' contrarian predisposition. The resulting mean-field approach exhibits
a chaotic behavior, which is our knowledge was rarely (if ever) observed before. 

What is remarkable is the appearance 
of coherent oscillations of the whole population also for the microscopic model, 
 in the presence of long-range connections (small-world). This may have important consequences for social scientists: the conflict between 
a liberal education (contrarian attitude) and the ever present social pressure may lead to unpredictable oscillation triggered by many quantities, 
in our widely connected world.

\section{Appendix}
\subsection{Continuous approximation and parameter mapping}\label{sec:scaling}
\begin{figure}
 \begin{center}
 \includegraphics[width=0.85\columnwidth]{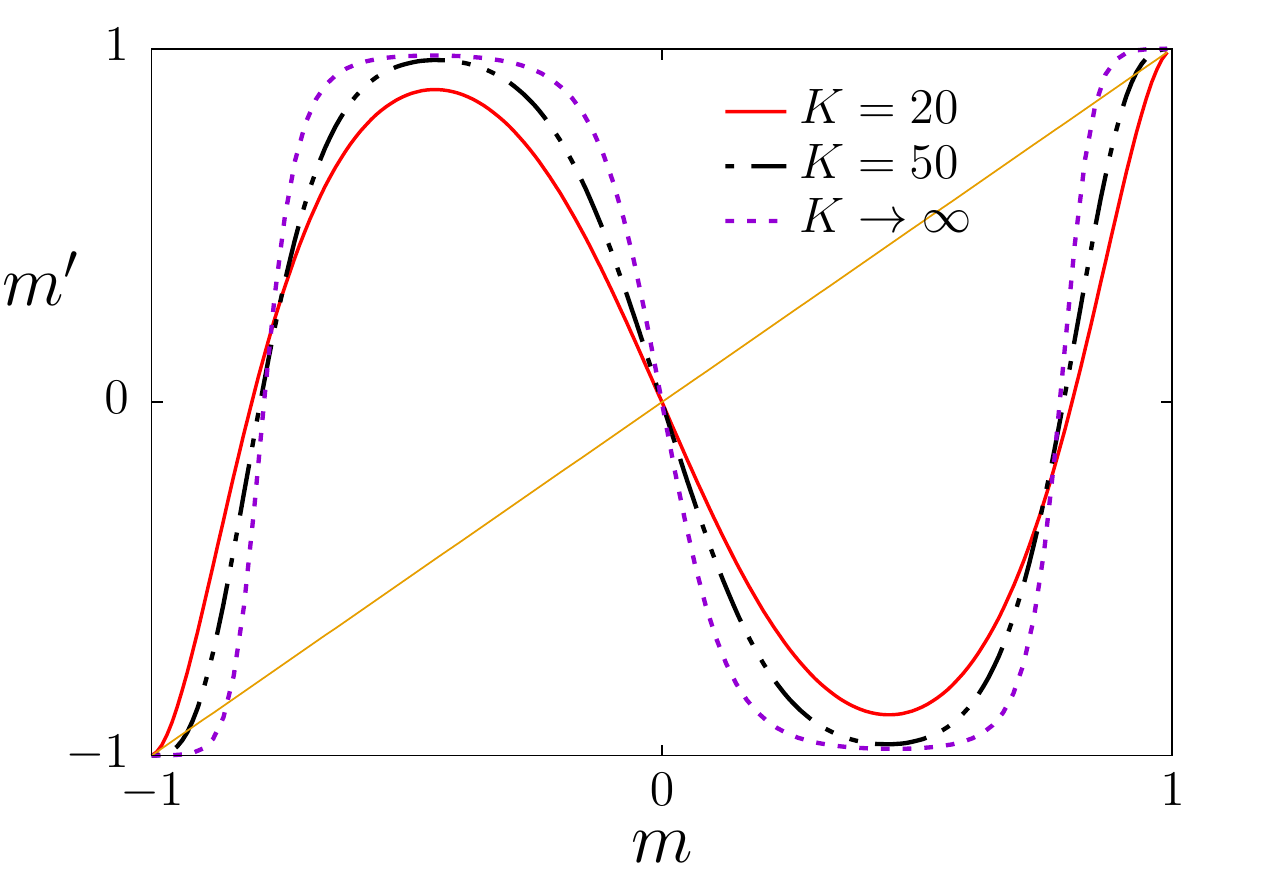}
 \end{center}
 \caption{\label{scaling} (Color online.) Scaling relation Eq.~\eqref{eq:scaling} for three values of the connectivity $K$, with $J(K=20)=-10$ and $W(K=20)=15$m], and the other values of $J$ and $W$ obtained from Eq.~\eqref{eq:scaling}.}
 \end{figure}

The similarities among bifurcation diagrams with different connectivity $K$ and coupling parameters $J$, $W$ can be explained by  using a  continuous approximation of the mean-field equation. 

By using Stirling's approximation for the binomial coefficients in Eq.~\eqref{eq:mf}, for small values of $m$~\cite{ott02}, we obtain
\begin{equation}\label{approx}
\begin{split}
	&\frac{1}{2^{K}}\binom{K}{k} (1+m)^{k}(1-m)^{K-k} \simeq\\
	&\qquad \sqrt{\frac{2K}{\pi (1-m^2)}} \exp\left[-\frac{2K}{1-m^2}\left(\frac{k}{K}-\frac{1+m}{2}\right)^2\right]
\end{split}
\end{equation}
and therefore, substituting $2k/K-1=x$, 
\begin{equation}\label{eq:continuous}
\begin{split}
m' =& \sqrt{\frac{K}{2\pi(1-m^2)}}\\
 & \times \int \limits_{-\infty}^{\infty} \mathrm{d}x\;\exp\left(-\frac{K(x-m)^2}{2(1-m^2)}\right)
  \tanh(Jx+Wx^3),
  \end{split}
\end{equation}
from which the convolution structure is evident. 

By developing $\tanh(-Jx+Wx^3)$ at first order (i.e., large $K$ and small values of $m$), we can compute the convolution, obtaining after remapping the first terms in powers of $m$, 
\begin{equation}\label{eq:tanh}
m' = \tanh(\tilde{J} m + \tilde{W} m^3)
\end{equation}
with 
\begin{equation}\label{eq:scaling}
	\begin{split}
		\tilde{J} &= J + \dfrac{3W}{K},\\
		\tilde{W} &= W\left(1-\dfrac{3}{K}\right).
	\end{split}
\end{equation}
Notice that in the limit $K\rightarrow\infty$, $\tilde J\rightarrow J$ and  $\tilde W\rightarrow W$. 

The relation between parameters $J,K$ and $J_1,K_1$ of two mean-field approximations  with different connectivities $K$ and $K_1$ is 
\begin{equation}\label{eq:correspondence}
\begin{split}
    J_1 &= J +\dfrac{3}{K}W\left(1-\dfrac{K-3}{K_1-3}\right),\\
	W_1 & =W \dfrac{K_1}{K}\dfrac{K-3}{K_1-3}.
\end{split}
\end{equation}

Since the approximation of the hyperbolic tangent is valid for small $x$, we expect that this scaling is better for large $K$, for which the convolution length is small. 
In Fig.~\ref{scaling} we report the scaling correspondence for some values of $K$. 

Since $J<0$ and $K>0$, the effect of this scaling is that of lowering the absolute value of $\tilde{J}$  and $\tilde{K}$ for small $K$ (larger than 3), so, given that for a large value of $K$ and certain values of  $J$ and $W$ the mean-field equation is chaotic, it may be reduced to a fixed point graph by lowering the connectivity $K$.

\section*{Acknowledgments} 
 This work was
partially supported by project PAPIIT-DGAPA-UNAM
IN109213. F.B. acknowledges partial financial support from European
Commission (FP7-ICT-2013-10) Proposal No. 611299 SciCafe 2.0.


\begin{thebibliography}{99}

\bibitem{Hegselmann}   R.  Hegselmann, U.  Krause,  \emph{Opinion  dynamics  and  bounded
confidence models, analysis, and simulation}, Journal of
Artificial Societies and
Social Simulation (JASSS) \textbf{5}, 1–33 (2002).

\bibitem{Deffuant} G. Deffuant, D. Neau, F. Amblard, and G. Weisbuch, \textit{Mixing beliefs among interacting agents}, 
  Adv. Complex Syst. {\bf 3} 87 (2000). doi:10.1142/S0219525900000078
%
\bibitem{review} C. Castellano, S. Fortunato and V. Loreto, \textit{Statistical physics of social dynamics},
  Rev. Mod. Phys. \textbf{81}, 591 (2009).

\bibitem{Stauffer} D. Stauffer, \textit{Sociophysics simulations II: opinion dynamics}, in 
\textit{Modelling Cooperative Behavior in the Social Sciences}, AIP Conf. Proc. \textbf{779}, 56 (2005). doi:10.1063/1.2008591

\bibitem{GalamReview} S. Galam,  \textit{Sociophysics: A review of Galam models}, 
International Journal of Modern Physics C \textbf{19}, 409 (2008). doi:10.1142/S0129183108012297

\bibitem{Galam1} S. Galam, \textit{Sociophysics: A Physicist's Modeling of Psycho-Political Phenomena}, (Springer, 2012). 
doi:10.1007/978-1-4614-2032-3 

\bibitem{guazzini} F. Bagnoli, T. Carletti, D. Fanelli, A. Guarino, A. Guazzini, \emph{Dynamical affinity in opinion dynamics modeling}, 
Phys. Rev. E \textbf{76}, 066105 (2007). doi:10.1103/PhysRevE.76.066105

\bibitem{BagnoliGuazziniLio}
F. Bagnoli, A. Guazzini, P. Li{\`o}, \emph{Human heuristics for autonomous agents}, Bio-Inspired Computing and Communication LNCS 5151 (Springer 2008) p. 340. doi:10.1007/978-3-540-92191-2\_30

\bibitem{GrottoGuazziniBagnoli}  R. Lauro Grotto, A. Guazzini, F. Bagnoli,
\emph{Metastable structures and size effects in small group dynamics}, 
Frontiers in psychology \textbf{5}, 699 (2014). doi:10.3389/fpsyg.2014.00699

\bibitem{GuazziniCiniBagnoliRamasco}
  A. Guazzini, A. Cini, F. Bagnoli, J.J. Ramasco, 
\emph{Opinion dynamics within a virtual small group: the stubbornness effect}, 
Frontiers in physics \textbf{3}, 65 (2015). doi:10.3389/fphy.2015.00065

\bibitem{ViloneCarlettibagnoliGuazzini}
D. Vilone, T. Carletti, F. Bagnoli, A. Guazzini,  \emph{The Peace Mediator effect: 
Heterogeneous agents can foster consensus in continuous opinion models},
J. Phys. A \textbf{462}, 84 (2016). doi:10.1016/j.physa.2016.06.082

\bibitem{latane} 
M. Lewenstein, A. Nowak, B. Latane, \textit{Statistical mechanics of social impact}
Phys. Rev. A \textbf{45}, 763 (1992). doi:10.1103/PhysRevA.45.763

\bibitem{masuda2013} N. Masuda, \textit{Voter models with contrarian agents}, Phys. Rev. E {\bf 88} 052803 (2013). 
doi:10.1103/PhysRevE.88.052803


\bibitem{crokidakis2014} N. Crokidakis, V. H. Blanco, and C. Anteneodo, \textit{Impact of contrarians and intransigents in a kinetic model of opinion dynamics},
  Phys. Rev. E {\bf 89} 013310 (2014). doi:10.1103/PhysRevE.89.013310
%
\bibitem{Independence} P. Nyczka,  K. Sznajd-Weron
\emph{Anticonformity or Independence?—Insights from Statistical Physics}, 
J. Stat. Phys. \textbf{151}, 174–202 (2013). doi:10.1007/s10955-013-0701-4

\bibitem{schneider2004} J.J Schneider, \textit{The influence of contrarians and opportunists on the stability of a democracy in the Sznajd model}, Int. J. Mod. Phys. C, {\bf 15} 659 (2004). doi:10.1142/S012918310400611X

\bibitem{delalama2005} M. S. de la Lama, J. M. L\'opez, and H. S. Wio, \textit{Spontaneous emergence of contrarian-like behaviour in an opinion spreading model},
  Europhys. Lett. {\bf 72} 851 (2005). doi:10.1209/epl/i2005-10299-3
  
\bibitem{corcos02} A. Corcos, J.-P. Eckmann, A. Malaspinas, Y. Malevergne, 
 D. Sornette, \textit{Imitation and contrarian behaviour: hyperbolic bubbles, crashes and chaos}, Quantitative Finance {\bf 2} 264 (2002).
 doi:10.1088/1469-7688/2/4/303
 
 
\bibitem{galam04} S. Galam,  \textit{Contrarian deterministic effects on opinion dynamics: ``the hung elections scenario''}, Physica A, {\bf 333} 453 (2004).
doi:10.1016/j.physa.2003.10.041

\bibitem{Biswas} S. Biswas, A. Chatterjee, and P. Sen, 
\textit{Disorder induced phase transition in kinetic models of opinion dynamics}, 
Physica A \textbf{391}, 3257 (2012). doi:10.1016/j.physa.2012.01.046


\bibitem{Galam-Gemrev} S. Galam, \textit{Modeling the Forming of Public Opinion: An approach from Sociophysics}, Global Economics and Management Review
  \textbf{18} 2 (2013). doi:10.1016/S2340-1540(13)70002-1


\bibitem{Galam-chaotic} C. Borghesi and S. Galam, \textit{Chaotic, staggered, and polarized dynamics in opinion forming: The contrarian effect}, Phys. Rev. E
  \textbf{73}, 066118 (2006). doi:10.1103/PhysRevE.73.066118

 \bibitem{sudoyi2013} S. D. Yi, S.K. Baek, C.P. Zhu, B.J. Kim, \textit{Phase transition in a coevolving network of conformist and contrarian voters},
  Phys. Rev. E {\bf 87} 012806 (2013). doi:10.1103/PhysRevE.87.012806
  
\bibitem{bagnoli2013} F. Bagnoli and R. Rechtman, \emph{Topological bifurcations in a model society of reasonable contrarians}, Phys. Rev. E, {\bf 88}
 062914 (2013). doi:10.1103/PhysRevE.88.062914

\bibitem{bagnoli2015} F. Bagnoli, R. Rechtman, 
\textit{Bifurcations in models of a society of reasonable contrarians and conformists},
  Phys. Rev. E \textbf{92}, 042913 (2015). doi:10.1103/PhysRevE.92.042913



\bibitem{Asch} S. E. Asch, Effects of group pressure on the
  modification and distortion of judgments, in {\em Groups,
  Leadership and Men}, H. Guetzkow editor, Carnegie Press, Pittsburgh
  PA (1951), pp. 177–190. S. E. Asch, {\em Social Psychology},
  Prentice Hall, Englewood Cliffs NJ, (1952). S. E. Asch,
  Psychological Monographs, \textbf{70}d, 1 (1956).
  
\bibitem{biswas2009} S. Biswas, P. Sen, \textit{Model of binary opinion dynamics: Coarsening and effect of disorder},
Phys. rev. E \textbf{80}, 027101 (2009). doi:10.1103/PhysRevE.80.027101

\bibitem{biswas2011a} S. Biswas, P. Sen, P. Ray, \textit{Opinion dynamics model with domain size dependent dynamics: 
novel features and new universality class}, Journal of Physics: Conference Series \textbf{297} (2011) 012003.
 doi:10.1088/1742-6596/297/1/012003

\bibitem{WattsStrogatz} D.J. Watts and S.H. Strogatz,  \emph{Collective dynamics of 'small-world' networks}, Nature 
 \textbf{393}, 409 (1998). doi:10.1038/30918

    \bibitem{Barabasi} L. Barab\'asi and R. Albert,  \emph{Emergence of scaling in random networks}, Science {\bf 286} 509
  (1999). doi:10.1126/science.286.5439.509
 
\bibitem{Klemm} K. Klemm, V. M. Eguiluz, R. Toral and M. San Miguel,
Phys. Rev. E \textbf{67}, 026120 (2003). doi:10.1103/PhysRevE.67.026120

\bibitem{WuZhou} B. Wu, D. Zhou, and L. Wang, \textit{Evolutionary dynamics on stochastic evolving networks for multiple-strategy games}, Phys. Rev. E. \textbf{84}, 046111 (2011). doi:10.1103/PhysRevE.84.046111

\bibitem{HolmeNewman}  P. Holme and M. E. J. Newman, 
\textit{Nonequilibrium phase transition in the coevolution of networks and opinions}, 
Phys. Rev. E \textbf{74}, 056108 (2006). doi:10.1103/PhysRevE.74.056108

\bibitem{barre} J.  Barré, A. Ciani, D. Fanelli, F. Bagnoli, S. Ruffo, \emph{Finite size effects for the Ising model on random graphs with varying dilution},
Physica A: Statistical Mechanics and its Applications \textbf{388}, 3413 (2009). doi:10.1016/j.physa.2009.04.024

\bibitem{goswami2011} S. Goswami, S. Biswas, P. Sen, \textit{Complex networks: Effect of subtle changes in nature of randomness}, 
Physica A: Statistical Mechanics and its Applications \textbf{390}, 972 (2011). doi:10.1016/j.physa.2010.10.024

\bibitem{biswas2011} S. Biswas, P. Sen, \textit{Effect of the nature of randomness on quenching dynamics of the Ising model on complex networks}, 
Phys. Rev. E  \textbf{84}, 066107 (2011). doi:10.1103/PhysRevE.84.066107



\bibitem{hipster} J. Touboul, \textit{The hipster effect: When anticonformists all look the same}, arXiv:1410.800,
\verb+http://arxiv.org/abs/1410.8001v1+
    
\bibitem{Derrida} B. Derrida, \textit{Dynamical Phase Transitions in in Spin Models and Automata} in H Beijeren (ed.), \textit{Fundamental Problems in Statistical Mechanics VII},  pp. 276, (Elsevier 1990).

\bibitem{NewmannDerrida} A.U. Neumann, B. Derrida,\textit{  Finite size scaling study of dynamical phase transitions in two dimensional models : Ferromagnet, symmetric and non symmetric spin glasses}, Journal de Physique \textbf{49}, (1988).     
doi:10.1051/jphys:0198800490100164700
    
\bibitem{Cirillo} E. N. M. Cirillo, F.R. Nardi, A. D. Polosa, \textit{Magnetic order in the Ising model with parallel dynamics}, Phys. Rev. E \textbf{64}, 057103 (2001). doi:10.1103/PhysRevE.64.057103

\bibitem{minority} W. B. Arthur,  \textit{Inductive Reasoning and Bounded Rationality [The El Farol Problem]},
The American Economic Review
  \textbf{84}, 406 (1994). 
  
\bibitem{minority1}  D. Challet and Y.-C. Zhang, \textit{Emergence of cooperation and organization in an evolutionary game}, Physica A
  \textbf{246}, 407 (1997). 
  
\bibitem{minority2}  D. Challet and M. Marsili, Y.-C. Zhang,
  {\em Minority Games}, Oxford University Press (2005). 
  
 \bibitem{minority3} A. C. C. Coolen,
  {\em The Mathematical Theory of Minority Games}, Oxford
         University Press (2005).
  
\bibitem{bagnoli2005} F. Bagnoli, F. Franci, R. Rechtman, \emph{Phase transitions of extended-range probabilistic cellular automata with two absorbing states}, Phys. Rev. E \textbf{71}, 046108 (2005). doi:10.1103/PhysRevE.71.046108

\bibitem{Dodds} P. S. Dodds, K. D. Harris, and C. M. Danforth, \textit{Limited Imitation Contagion on Random Networks: Chaos, Universality, and Unpredictability},
  Phys. Rev. Let. \textbf{110}, 158701 (2013). doi:10.1103/PhysRevLett.110.158701

\bibitem{Harris}  K.D. Harris, C.M. Danforth, and  P.S. Dodds,  \textit{Dynamical influence processes on networks: General theory and applications to social contagion},
Phys. Rev. E \textbf{88}, 022816 (2013). doi:10.1103/PhysRevE.88.022816

\bibitem{Boltzmann} L. Boltzmann, {\em Vorlesungen \"uber Gastheorie},
  Leipzig, J. A. Barth, Part I, 1896, Part II, 1898. English
  translation by S. G.  Brush, {\em Lectures on Gas Theory},
  (University of California Press, 1964), Chapter I, Sec. 6.

\bibitem{Suzuki} M. Suzuki, \textit{Solution and Critical Behavior of Some ``Three-Dimensional'' Ising Models with a Four-Spin Interaction},
Phys. Rev. Lett. \textbf{28}, 507 (1972). doi:10.1103/PhysRevLett.28.507

\bibitem{Sherrington} C. Castelnovo, C. Chamon, D. Sherrington, \textit{Quantum mechanical and information theoretic view on classical glass transitions},
Phys. Rev. B \textbf{81}, 184303 (2010). doi:10.1103/PhysRevB.81.184303

\bibitem{SS-bif}  F. Bagnoli, T. Matteuzzi, R. Rechtman, \textit{Topological Phase Transitions in the Nonlinear Parallel Ising Model}, Acta Physica Polonica B Proc. Suppl. \textbf{9}, 37 (2016). doi:10.5506/APhysPolBSupp.9.37

\bibitem{SS-meta} F. Bagnoli, T. Matteuzzi, R. Rechtman, \textit{Phase Transitions and Metastable States in the Parallel Ising Model}, Acta Physica Polonica B Proc. Suppl. \textbf{9}, 25 (2016). doi:10.5506/APhysPolBSupp.9.25
 
\bibitem{ott02} E. Ott, textit{Chaos in Dynamical Systems} (Cambridge University Press, Cambridge UK, 2002). ISBN: 9780521010849
    
\bibitem{kolmogorov58} A.N. Kolmogorov, New Metric Invariant of
  Transitive Dynamical Systems and Endomorphisms of Lebesgue Spaces,
  Doklady of Russian Academy of Sciences {\bf 119}, N5, 861-864
  (1958).  
  
\bibitem{kolmogorov59} A.N. Kolmogorov, Entropy per unit time as a metric
  invariant of automorphism, Doklady of Russian Academy of Sciences
  {\bf 124}, 754-755 (1959).

\bibitem{sinai59} Ya.G. Sinai, On the Notion of Entropy of a Dynamical
  System, Doklady of Russian Academy of Sciences {\bf 124} 768-771 (1959).
  
\bibitem{strogatz} S.H. Strogatz, \textit{Nonlinear Dynamics And Chaos} (Westview Press, Boulder, CO 2015).  ISBN: 9780813349107


\end{thebibliography}
\end{document}